\RequirePackage{snapshot}
\documentclass[twocolumn,nofootinbib]{aastex62}
\usepackage{calc}
\usepackage{mathtools,graphicx}
\usepackage{pifont}
\usepackage{bm}
\usepackage{microtype}
\usepackage{booktabs}
\usepackage{times}
\usepackage[varg]{txfonts}
\usepackage[utf8]{inputenc}
\usepackage[caption=false]{subfig}
\usepackage{paralist}
\usepackage[normalem]{ulem}
\usepackage{multirow}
\usepackage{etoolbox}
\usepackage{longtable}
\usepackage{xstring}
\usepackage{xparse}
\usepackage[nolist,nohyperlinks,printonlyused]{acronym}
\usepackage{dcolumn}

\definecolor{LinkColor}{rgb}{0.75, 0, 0}
\definecolor{CiteColor}{rgb}{0, 0.5, 0.5}
\definecolor{UrlColor}{rgb}{0, 0, 0.75}
\hypersetup{linkcolor=LinkColor}
\hypersetup{citecolor=CiteColor}
\hypersetup{urlcolor=UrlColor}
\allowdisplaybreaks
\DeclareFontFamily{OT1}{pzc}{}
\DeclareFontShape{OT1}{pzc}{m}{it}{<-> s * [1.10] pzcmi7t}{}
\DeclareMathAlphabet{\mathpzc}{OT1}{pzc}{m}{it}

\setcitestyle{notesep={}}

\AtBeginDocument{
    \newwrite\bibnotes
    \def\bibnotesext{Notes.bib}
    \immediate\openout\bibnotes=\jobname\bibnotesext
    \immediate\write\bibnotes{@CONTROL{REVTEX41Control}}
    \immediate\write\bibnotes{@CONTROL{%
    apsrev41Control,author="08",editor="1",pages="0",title="0",year="1"}}
     \if@filesw
     \immediate\write\@auxout{\string\citation{apsrev41Control}}
    \fi
}

\newtoggle{fullauthorlist}
\toggletrue{fullauthorlist}

\newtoggle{endauthorlist}
\toggletrue{endauthorlist}

\graphicspath{{./fig/}}

\usepackage{scrextend}
\usepackage{textgreek}
\usepackage{xspace}

\usepackage{graphicx}
\usepackage{makerobust}
\makeatletter
\@ifundefined{caption@xref}{}{%
  \MakeRobustCommand\caption@xref
}
\makeatother

\def\mnras{\ref@jnl{MNRAS}}             

\renewcommand{\today}{\number\day\space\ifcase\month\or
  January\or February\or March\or April\or May\or June\or
  July\or August\or September\or October\or November\or December\fi
  \space\number\year}

\newtoggle{cleancomments}
\togglefalse{cleancomments}

\newcommand{\IMRXPHM}{\textsc{IMRPhenomXPHM}\xspace}

\newcommand{\gstlal}{\textsc{GstLAL}\xspace}
\newcommand{\pycbc}{\textsc{PyCBC}\xspace}
\newcommand{\cwb}{\textsc{cWB}\xspace}

\newcommand{\golum}{\textsc{golum}\xspace}

\newcommand{\gw}[1][]{gravitational wave#1 (GW#1)\renewcommand{\gw}[1][]{GW##1\xspace}\xspace}
\newcommand{\cbc}[1][]{compact binary coalescence#1 (CBC#1)\renewcommand{\cbc}[1][]{CBC##1\xspace}\xspace}
\newcommand{\bns}[1][]{binary neutron star#1 (BNS#1)\renewcommand{\bns}[1][]{BNS##1\xspace}\xspace}
\newcommand{\bbh}[1][]{binary black hole#1 (BBH#1)\renewcommand{\bbh}[1][]{BBH##1\xspace}\xspace}
\newcommand{\nsbh}[1][]{neutron star--black hole#1 (NSBH#1)\renewcommand{\nsbh}[1][]{NSBH##1\xspace}\xspace}
\newcommand{\pe}{parameter estimation (PE)\renewcommand{\pe}{PE\xspace}\xspace}
\newcommand{\pisn}{pair instability supernova (PISN)\renewcommand{\pisn}{PISN\xspace}\xspace}
\newcommand{\kde}[1][]{kernel density estimator#1 (KDE#1)\renewcommand{\kde}[1][]{KDE##1\xspace}\xspace}
\newcommand{\fap}[1][]{false-alarm probability#1 (FAP#1)\renewcommand{\fap}[1][]{FAP##1\xspace}\xspace}
\newcommand{\far}[1][]{false-alarm rate#1 (FAR#1)\renewcommand{\far}[1][]{FAR##1\xspace}\xspace}
\newcommand{\sgwb}{stochastic GW background (SGWB)\renewcommand{\sgwb}{SGWB\xspace}\xspace}
\newcommand{\snr}[1][]{signal-to-noise ratio#1 (SNR#1)\renewcommand{\snr}[1][]{SNR##1\xspace}\xspace}

\newcommand{\Msun}{\ensuremath{\text{M}_\odot}}
\newcommand{\Mc}{\ensuremath{\mathcal{M}}}

\newcommand{\cohratio}{\mathcal{C}^\mathrm{L}_\mathrm{U}}
\newcommand{\Clu}{\mathcal{C}^\mathrm{L}_\mathrm{U}}
\newcommand{\Blu}{\mathcal{B}^\mathrm{L}_\mathrm{U}}
\newcommand{\BLU}{\textsf{B}^{\rm L}_{\rm U}}

\newcommand{\BMLU}{\mathcal{B}_\mathrm{U}^\mathrm{Micro}}
\newcommand{\zL}{z_\mathrm{l}}
\newcommand{\MzL}{M^z_\mathrm{L}}

\def\fdm{{f_\mathrm{DM}}}
\def\LambdaL{{\Lambda_\ell}}
\def\NL{{N_\ell}}

\newcommand{\RateGalaxyDouble}{1.9$-$11.0 $\times 10^{-4}$  } 
\newcommand{\RateGalaxySingle}{5.0$-$19.5 $\times 10^{-4}$  }
\newcommand{\RateClusterDouble}{0.8$-$4.4 $\times 10^{-4}$ }
\newcommand{\RateClusterSingle}{2.0$-$7.6 $\times 10^{-4}$}

\begin{document}
\begin{acronym}
\acro{GW}{gravitational wave}
\acro{SGWB}{stochastic gravitational-wave background}
\acro{MBTA}{Multi-Band Template Analysis}
\acro{SIS}{singular isothermal sphere}
\acro{SIE}{singular isothermal ellipsoid}
\acro{FPP}{false-positive probability}
\acro{FAP}{false-alarm probability}
\acro{ML}{machine learning}
\acro{SNR}{signal-to-noise ratio}
\acro{PSD}{power spectral density}
\acro{BBH}{binary black hole}
\acro{BNS}{binary neutron star}
\acro{NSBH}{neutron star-black hole}
\acro{PE}{parameter estimation}
\acro{FAR}{false-alarm rate}
\acro{BH}{black hole}
\end{acronym}

\title{Search for gravitational-lensing signatures in the full third observing run of the LIGO--Virgo network}

\iftoggle{endauthorlist}{
 \let\mymaketitle\maketitle
 \let\myauthor\author
 \let\myaffiliation\affiliation
 \author{The LIGO Scientific Collaboration, the Virgo Collaboration and the KAGRA Collaboration}
}{
 \iftoggle{fullauthorlist}{
	 \author{R.~Abbott}
\affiliation{LIGO Laboratory, California Institute of Technology, Pasadena, CA 91125, USA}
\author{H.~Abe}
\affiliation{Graduate School of Science, Tokyo Institute of Technology, Meguro-ku, Tokyo 152-8551, Japan  }
\author{F.~Acernese}
\affiliation{Dipartimento di Farmacia, Universit\`a di Salerno, I-84084 Fisciano, Salerno, Italy  }
\affiliation{INFN, Sezione di Napoli, I-80126 Napoli, Italy  }
\author[0000-0002-8648-0767]{K.~Ackley}
\affiliation{OzGrav, School of Physics \& Astronomy, Monash University, Clayton 3800, Victoria, Australia}
\author{S.~Adhicary}
\affiliation{The Pennsylvania State University, University Park, PA 16802, USA}
\author[0000-0002-4559-8427]{N.~Adhikari}
\affiliation{University of Wisconsin-Milwaukee, Milwaukee, WI 53201, USA}
\author[0000-0002-5731-5076]{R.~X.~Adhikari}
\affiliation{LIGO Laboratory, California Institute of Technology, Pasadena, CA 91125, USA}
\author{V.~K.~Adkins}
\affiliation{Louisiana State University, Baton Rouge, LA 70803, USA}
\author{V.~B.~Adya}
\affiliation{OzGrav, Australian National University, Canberra, Australian Capital Territory 0200, Australia}
\author{C.~Affeldt}
\affiliation{Max Planck Institute for Gravitational Physics (Albert Einstein Institute), D-30167 Hannover, Germany}
\affiliation{Leibniz Universit\"at Hannover, D-30167 Hannover, Germany}
% ...

 }{
 \author{The LIGO Scientific Collaboration, the Virgo Collaboration and the KAGRA Collaboration}
}
}

\begin{abstract}
\noindent
Gravitational lensing by massive objects along the line of sight to the source causes distortions of gravitational wave-signals; such distortions may reveal information about fundamental physics, cosmology and astrophysics.
In this work, we have extended the search for lensing signatures to all binary black hole events from the third observing run of the LIGO--Virgo network.
We search for repeated signals from strong lensing by 1) performing targeted searches for subthreshold signals, 2) calculating the degree of overlap amongst the intrinsic parameters and sky location of pairs of signals, 3) comparing the similarities of the spectrograms amongst pairs of signals, and 4) performing dual-signal Bayesian analysis that takes into account selection effects and astrophysical knowledge.
We also search for distortions to the gravitational waveform caused by 1) frequency-independent phase shifts in strongly lensed images, and 2) frequency-dependent modulation of the amplitude and phase due to point masses.
None of these searches yields significant evidence for lensing.
Finally, we use the non-detection of gravitational-wave lensing to constrain the lensing rate based on the latest merger-rate estimates and the fraction of dark matter composed of compact objects.

\end{abstract}

\section{Introduction}
\label{sec:intro}
Gravitational lensing occurs when a massive object bends spacetime in a way that alters the path or properties of a propagating wave.
Gravitational lensing is expected to affect \acp{GW}, resulting, for example, in repeated signals, (de-)magnification of the amplitude, phase shifts, and beating patterns~\citep{Ohanian:1974ys,Thorne:1982cv,Deguchi:1986zz,Wang:1996as,Nakamura:1997sw,Takahashi:2003ix}.
The exact alteration of the gravitational waveform depends on the nature of the lens system.

For massive lenses, gravitational lensing changes the \ac{GW} amplitude without affecting the frequency evolution~\citep{Wang:1996as,Dai:2017huk,Ezquiaga:2020gdt}.
Moreover, such systems may also produce multiple signals observed as repeated events separated by a time delay of minutes to months for galaxies~\citep{Ng:2017yiu,Li:2018prc,Oguri:2018muv}, and up to years for galaxy clusters~\citep{Smith:2017mqu,Smith:2018gle,Smith:2019dis,Robertson:2020mfh,Ryczanowski:2020mlt}.
The current-generation \ac{GW} detector network has a realistic chance of detecting the first lensed signal within its operation period~\citep{Ng:2017yiu,Li:2018prc,Oguri:2018muv}.

For low-mass lenses, such as stars or compact objects, microlensing introduces beating patterns in the waveform~\citep{Deguchi:1986zz,Nakamura:1997sw,Takahashi:2003ix,Cao:2014oaa,Jung:2017flg,Lai:2018rto,Christian:2018vsi,Dai:2018enj,Diego:2019rzc}.
More generally, a field of light lenses may produce even more complex patterns on the gravitational waveform \citep{Diego:2019lcd,Pagano:2020rwj,Cheung:2020okf}.
Under the right conditions and with sufficient knowledge about the lens, these beating patterns may be observable with current-generation \ac{GW} detectors.

The detection of lensed \acp{GW} paves the way for numerous scientific pursuits, including source localization~\citep{Hannuksela:2020xor} and characterization~\citep{Lai:2018rto,Diego:2019rzc,Oguri:2020ldf}, precision cosmology~\citep{Sereno:2011ty,Liao:2017ioi,Cao:2019kgn,Li:2019rns,Hannuksela:2020xor}, and tests of general relativity~\citep{Baker:2016reh,Collett:2016dey,Fan:2016swi,Goyal:2020bkm,Ezquiaga:2020dao}.
Indeed, the prospects for fundamental physics and astrophysics have sparked a wide interest in searching for lensed \acp{GW}.
Previous work from the LIGO--Virgo Collaboration has considered a range of strong and microlensing signatures for events in the first half of the third observing run (O3a) \citep{LIGOScientific:2021izm}.
Nevertheless, these studies have yielded no confident evidence for \ac{GW} lensing.

In this work, we search for a variety of lensing signatures in the third LIGO Scientific, Virgo, and KAGRA (LVK) Collaboration Gravitational-Wave Transient Catalog (GWTC-3)~\citep{LIGOScientific:2021djp} and study its implications for \ac{GW} lensing.
In particular, we expand on the lensing results presented for the first half of the third observing run of the LIGO--Virgo network (O3a)~\citep{LIGOScientific:2021izm} by including the signals found in the second half of the third observing run (O3b) and by including additional analyses to further test the lensing hypothesis and interpret their outcomes.
First, we search for the effects of strong lensing by studying the similarity and lensing evidence for pairs of \ac{BBH} mergers.
We consider both pairs of detected mergers (super-threshold) and pairs formed by detected mergers and candidates that nominally fall below the detection threshold (sub-threshold) with consistent waveform morphologies.
Second, we search for evidence of microlensing induced by point-mass lenses.
Finally, we constrain the expected rate of lensed signals, \ac{BH} merger-rate density, and the fraction of dark matter composed of compact objects.

It is important to note that GWTC-3 is a cumulative catalog describing all the \ac{GW} transients found in all observing runs to date: O1, O2, O3a, and O3b.  O1 made observations between 2015 September 12 00:00 UTC to 2016 January 19 16:00 UTC, O2 between 2016 November 30 16:00 UTC to 2017 August 25 22:00 UTC, O3a between 2019 April 1 15:00 UTC to 2019 October 1 15:00 UTC, and O3b between 2019 November 1 15:00 UTC to 2020 March 27 17:00 UTC.

Results of all analyses in this paper and associated data products can be found in \citet{datarelease}.
\ac{GW} strain data \citep{gwosc:gwtc3} and posterior samples \citep{ligo_scientific_collaboration_and_virgo_2021_5546663} for all events from GWTC-3 are available from the Zenodo platform or the Gravitational Wave Open Science Center~\citep{Abbott:2019ebz}.

\section{Data and Events}
\label{sec:data}
The analyses presented here expand on the lensing results from the first half of O3 (also referred to as O3a) by documenting new results from the second half of O3 (also referred to as O3b) using GWTC-3~\citep{GWTC3}.
The O3a lensing results paper~\citep{LIGOScientific:2021izm} used the GWTC-2 catalog~\citep{GWTC2}.
Since then, GWTC-2.1~\citep{GWTC-2.1} has reclassified 2 of the candidates used in the O3a lensing paper as having a probability of astrophysical origin of less than 0.5 and are not included in the results described here, specifically GW190424\_180648 and GW190909\_114149.
GWTC-3 also includes 5 events that were identified by the O3a lensing sub-threshold counterpart image search, namely GW190925\_233845, GW190426\_190642, GW190725\_184728, GW190805\_211137, and GW190916\_200658.

Various instrumental upgrades have led to more sensitive data in O3b,
with a median \ac{BNS} inspiral ranges~\citep{FinnChernoff:1993,PhysRevD.85.122006} of 115 Mpc in O3b compared to 108 Mpc in O3a for LIGO Hanford, 133 Mpc in O3b compared to 135 Mpc O3a for LIGO Livingston, and 51 Mpc in O3b compared to 45 Mpc in O3a for Virgo~\citep{GWTC3}.
The duty factor for at least one detector being online was 96.6\%; for any two detectors being online at the same time was 85.3\%; and for all three detectors together was 51\%.
Further details regarding instrument performance and data quality for O3b are available in \citet{GWTC3,Davis2021,Acernese:2022jes}.

The LIGO and Virgo detectors used a photon recoil-based calibration~\citep{Karki:2016pht,Cahillane:2017vkb,Viets:2017yvy} resulting in a complex-valued, frequency-dependent detector response.
Previous studies have documented the systematic error and uncertainty bounds for O3b strain calibration in LIGO~\citep{Sun_2020,Sun2021} and Virgo~\citep{Acernese2021}. 

Transient noise sources, referred to as glitches, contaminate the data and can affect the confidence of candidate detections.
Times affected by glitches and other data quality issues are identified so that searches for \ac{GW} events can exclude (veto) these periods of poor data quality~\citep{TheLIGOScientific:2016zmo,LIGOScientific:2019hgc,Davis:2021ecd,nguyen2021environmental,Fiori2020}.
In addition, several known persistent noise sources are subtracted from the data using information from witness auxiliary sensors~\citep{Driggers:2018gii,Davis:2018yrz}.

Candidate events, including those reported in~\citet{GWTC3} and the new candidates found by the search for sub-threshold counterpart images in Sec.~\ref{sec:subthreshold} of this paper, have undergone a validation process to evaluate if instrumental artifacts could affect the analysis; this process is described in detail in Sec. 5.5 of~\citet{Davis:2021ecd}.
This process can also identify data quality issues that need further mitigation for individual events, such as the subtraction of glitches~\citep{PhysRevD.103.044006, Davis_glitchSub22} and non-stationary noise couplings~\citep{Vajente:2019ycy}, before executing \ac{PE} algorithms.
See Table XIV of \citet{GWTC3} for the list of events requiring such mitigation.

The GWTC-3 catalog~\citep{GWTC3} contains 35 events from O3b in addition to the 55 previous events from previous observing runs~\citep{GWTC-2.1} with a \ac{FAR} below two per year, and an expected rate of contamination from detector noise less than 10--15\%~\citep{GWTC3}. 
We neglect the potential contamination in this analysis.
These events were identified by four search pipelines: one minimally modeled transient search \cwb~\citep[][]{Klimenko:2004qh,Klimenko:2005xv,Klimenko:2006rh,Klimenko:2011hz,Klimenko:2015ypf} and three matched-filter searches \gstlal~\citep{Sachdev:2019vvd, Hanna:2019ezx, Messick:2016aqy}, \ac{MBTA}~\citep{Adams_2016, Aubin_2021}, and \pycbc~\citep{Allen:2005fk, Allen:2004gu, Canton:2014ena, Usman:2015kfa,Nitz:2017svb}.
Their parameters were estimated through Bayesian inference using the \texttt{bilby}~\citep{Ashton:2018jfp, Smith:2019ucc,Romero-Shaw:2020owr} and \texttt{RIFT}~\citep{PhysRevD.92.023002, PhysRevD.96.104041, PhysRevD.99.084026} packages.
Both the matched-filter searches and \ac{PE} use a variety of \ac{BBH} waveform models which generally combine knowledge from post-Newtonian theory, the effective-one-body formalism, and numerical relativity \citep[for general introductions to these approaches, see][ and references therein]{Blanchet:2013haa,Damour:2016bks,Palenzuela:2020tga,Schmidt:2020ekt}.
The analyses in this paper rely on the same methods, and the specific waveform models and analysis packages used are described in each section.

Of the 35 events from O3b, 31 are likely \acp{BBH}, while four have component masses consistent with being below 3\,$\Msun$~\citep{LIGOScientific:2021qlt, GWTC3}, thus potentially containing a neutron star.
We consider these 35 events in the analyses documented in this paper.
Specifically, we use the following input data sets for each analysis.
The searches for sub-threshold counterpart images in Sec.~\ref{sec:subthreshold} cover the whole O3 strain data set, using the same data quality veto choices as in~\citet{GWTC3} but a strain data set consistent with the \ac{PE} analyses: the final calibration version of LIGO data~\citep{Sun2021} with additional noise subtraction~\citep{Vajente:2019ycy}.
The posterior-overlap analysis in Sec.~\ref{sec:posterioroverlap} starts from the posterior samples released with GWTC-3~\citep{gwosc:gwtc3}.
The joint-\ac{PE} analyses in Sec.~\ref{sec:jointpe} and microlensing analysis in Sec.~\ref{sec:microlensing} reanalyze the strain data in short segments around the event times, available from the same data release, with data selection and noise mitigation choices matching those of the \ac{PE} analyses in~\citet{GWTC3}.

\section{Strong lensing}
\label{sec:multiimages}
If a \ac{GW} travels close enough to a massive lens, it will produce multiple images, with the number of images depending on the lens profile and source lens geometry. 
This regime is known as the strong-lensing limit. 
Each of these lensed images $h_j^\mathrm{L}$ will have a change in its amplitude, arrival time and phase compared to the emitted signal $h$~\citep{1992grle.book.....S}: 
\begin{equation}
h_j^\mathrm{L}(f) = \sqrt{|\mu_j|}\exp\left[{i2\pi f\Delta t_j-i\mathrm{sign}(f)n_j\pi}\right]h(f)\,,
\label{eq:strong-lens}
\end{equation}
for $n_j=0,1/2,1$ for type I, II and III images, which correspond to different minima of the lensing potential. 
While the magnification $\mu_j$ and time delay $\Delta t_j$ do not affect the waveform morphology (they are completely degenerate with the luminosity distance and coalescence time) the frequency-independent lensing phase shift $n_j\pi$ could induce distortions when the signal has multiple frequency components~\citep{Dai:2017huk,Ezquiaga:2020gdt}. 
In particular, this occurs for type II images since type I does not have a phase shift, and type III only flips the overall sign, which is degenerate with shifting the polarization angle by $\pi/2$. 
The $\mathrm{sign}(f)$ term is only there to ensure that the time domain waveform is real.  

Making a distinction between effects that do and do not change the waveform morphology, we divide our search into two parts.
First, we search for pairs of events consistent with the strong-lensing hypothesis. 
Some of these pairs will have sufficiently strong amplitudes that can be identified as confident detections (super-threshold) by the search pipelines used in~\citet{GWTC2,GWTC-2.1,GWTC3}, while others may have not been identified as signals (sub-threshold) because of the relative de-magnification. 
Our searches will include both sub- and super-threshold pairs. 
A pair is the minimum association, but higher multiplicities are also possible. 
Then, we search for strong lensing focusing on the distortion of type II images.

    \subsection{Sub-threshold Search}
    \label{sec:subthreshold}
    In this section, we describe the search for possible sub-threshold counterparts of super-threshold detections from O3.
We perform searches over all O3 strain data following the rules for data selection described in \citet{GWTC-2.1} and \citet{GWTC3}.
A general search for \acp{GW} uses a large template bank covering a broad parameter space as we have no prior information about the signal subspace, resulting in a high trials factor and hence incurring a high noise background.
Sub-threshold (lensed) \acp{GW} with smaller amplitudes will therefore be easily buried in the noise without being identified as detections as they cannot pass the usual detection threshold.

To uncover these sub-threshold (lensed) signals, we have to effectively reduce the noise background while keeping the targeted foreground (i.e. the signals) constant~\citep{Li:2019osa,McIsaac:2019use,Dai:2020tpj}. 
The strong lensing hypothesis asserts that lensed \acp{GW}, super-threshold or sub-threshold, coming from the same origin have identical waveforms apart from an overall scaling factor and a Morse phase factor as described in Eq.~\ref{eq:strong-lens}, and hence should have consistent inferred intrinsic masses and spins.
\footnote{
The Morse phase factor for different image types has not been considered in the search described here.
Should a \ac{GW} include detectable higher-order multipole moments, then the Morse phase factor will cause complicated changes to the waveforms, inducing a loss in the search sensitivity.
}
Therefore, we can construct a reduced template bank with only templates that have masses and spins similar to those of a target super-threshold detection.
Using the reduced bank lowers the trials factors and noise background and effectively searches for previously unidentified possible sub-threshold lensed counterparts to the target detection. 
For each known candidate from O3 with a probability of astrophysical origin $p_\text{astro}>0.5$, we create a reduced template bank using their respective public posterior mass and spin samples released with GWTC-3~\citep{gwosc:gwtc3}, ensuring that the templates will match well with their respective target events while improving the ranking statistics of the search for similar events, and hence potentially returning new candidates that previously did not reach the threshold $p_\text{astro}>0.5$ in GWTC-3.
Details of how the reduced banks are constructed can be found in~\citet{Li:2019osa}.

Given these template banks, we proceed with configurations and procedures as outlined in \citet{LIGOScientific:2021izm} to produce a priority list of potential lensed candidates matching each target event, using \gstlal~\citep{gstlal-methods,Sachdev:2019vvd} as the search pipeline.
The list of candidates obtained is again further vetted using a sky location consistency check detailed in~\citet{Wong:2021lxf,LIGOScientific:2021izm} to ensure the candidates have consistent sky location with the target event.
To avoid false dismissal at this step, we only veto candidates with an overlap in $90\%$ credible region of the sky location \mbox{$O_{\rm 90\% CR}=0$}.
All candidates with non-vanishing localization overlap are kept for further follow-up with data quality checks as discussed in Sec.~\ref{sec:data}.

In Table~\ref{table:subthresh-search-results-top}, we list the top five candidates from the \emph{individual} targeted searches for counterparts of the detections reported in O3.
As in the O3a lensing paper \citep{LIGOScientific:2021izm}, we do not assess in detail the probability of astrophysical origin for each of these.
It is also important to note that the reported \far[s] do not indicate how likely each trigger is a lensed counterpart of the target event, but only how likely noise produces a trigger with a ranking statistic higher or equal to that of the candidate under consideration using these reduced template banks. 
Similar to \citet{LIGOScientific:2021izm}, we account for the fact that we have analyzed $\sim 332$ days of data multiple times for a total of $76$ events, and set the \ac{FAR} threshold to be $1$ in $69$ years (i.e. $4.59\times10^{-10}$ Hz).
We followed up on the top two candidates listed that passed the \ac{FAR} threshold through \texttt{golum}'s joint PE analysis \citep[, discussed in \ref{sec:jointpe}]{Janquart:2021qov}. The results are included in Table~\ref{table:subthresh-search-results-top}. Since both pairs of candidates have mildly \emph{negative} $\log_{10}$ coherence ratios, showing that there is no evidence supporting the lensing hypothesis for either of these pairs, we did not further follow them up with the more computationally intensive \texttt{hanabi} analysis \citep[, discussed in \ref{sec:jointpe}]{Lo:2021nae}.

\begin{table*}[t]
\caption{\label{table:subthresh-search-results-top}
Top $5$ candidates from individual sub-threshold searches for strongly-lensed counterpart images of O3 events from GWTC-3.
}
\hspace{-1.75cm}
\resizebox{1.1\textwidth}{!}{
\begin{tabular}{cccccccc}
\tableline \tableline
Target event & Lensed candidate (UTC) & $\Delta t$ [days] & $(1+z)\Mc (M_\odot)$  & $(1+z)\Mc^\text{target} (M_\odot)$ & \far $\left[\text{yr}^{-1}\right]$ & $O_{\rm 90\% CR}$ [\%] & $\log_{10}(\Clu)$\\
\tableline
GW$190930\_133541$ & 19-08-05 13:43:48 & $-56.0$ & 10.2 & $9.86$ & 0.002 & $61.00\%$ & $-6.4$\\
GW$191204\_171526$ & 19-08-05 13:43:48 & $-121.1$ & 10.2 & $9.66$ & 0.006 & $25.40\%$ & $-12.2$\\
GW$190828\_065509$ & 19-11-12 12:13:18 & $76.2$ & 45.5 & $17.3$ & 0.023 & $18.00\%$ & - \\
GW$190725\_174728$ & 19-08-05 13:43:48 & $10.8$ & 10.2 & $8.88$ & 0.038 & $39.50\%$ & -\\
GW$190828\_065509$ & 20-02-06 07:24:59 & $162.0$ & 15.1 & $17.3$ & 0.154 & $36.00\%$ & - \\
\tableline
\end{tabular}
} 
\tablecomments{
The first column lists the target event from O3. 
The second column shows the time (YY-MM-DD HH-MM-SS) in UTC of the found sub-threshold candidate.
The third column shows the time difference (in days) between the candidate and the target event.
The fourth column shows the redshifted chirp mass of the template that found the trigger.
The fifth column shows the redshifted chirp mass of the target event.
The sixth column shows the \far[s] from the individual search for the new candidate from the second column.
The seventh column shows the percentage overlap of the $90\%$ sky localization regions between the candidate and the target event.
The eighth column shows the $\log_{10}$ coherence ratio obtained from \texttt{golum}'s joint PE analysis.
}
\end{table*}

    \subsection{Preliminary Identification of Lensed Pair Candidates}
    \label{sec:posterioroverlap}
    Multiple, non-overlapping images produced by strongly lensed \ac{GW} signals have identical phase evolution, and therefore their intrinsic parameters (as well as their orbit's inclination with respect to the line of sight) are expected to have overlapping posteriors.  
In addition, the angular separations of images (produced by galaxies or galaxy clusters) are several orders of magnitude smaller than the uncertainties associated with their \ac{GW} sky location. 
As a result, their sky localisations will also overlap. (As in the previous section, the Morse phase for different image types is not considered here.)

Under these assumptions, a Bayes-factor statistic ($\BLU$) that assesses the consistency between a lensed candidate pair's posterior distributions of intrinsic parameters, sky location, and inclination angle (and thus acts as a discriminator between the lensed and unlensed hypotheses) can be constructed \citep{Haris:2018vmn}. 
To convert this statistic to a \ac{FPP},%
\footnote{
	\ac{FAR} and \ac{FPP}, while conceptually similar, pertain to different contexts in this work.
	In particular, we use \ac{FPP} exclusively for significances associated with candidate lensed pairs to discriminate them from unlensed pairs.
	On the other hand, a \ac{FAR} is associated with the significance assigned to individual candidate \ac{GW} signal events.
} a background distribution of unlensed $\BLU$ needs to be estimated. 

To that end, we conduct an injection campaign involving \acp{BBH} only, in which we sample component masses $m_{1,2}$ from a power-law distribution \citep{Abbott:2016nhf} in the range $(10$--$50 M_{\odot})$.
We assume that the redshift distribution of \acp{BBH} is similar to population synthesis simulations of isolated binary evolution  \citep{Belczynski:2005mr,Belczynski:2010tb,Dominik:2013tma,Eldridge:2018nop, Bouffanais:2021wcr, Zevin:2020gbd}.
All other parameters are sampled from uninformative prior distributions \citep{Haris:2018vmn}.
We inject the simulated signals into Gaussian noise with O3a representative \ac{PSD} for a LIGO--Virgo detector network.  
We compute $\BLU~$for all possible pairs in this injection set, and following \citet{LIGOScientific:2021izm}, we assign an \ac{FPP} to a candidate pair using its $\BLU$.
Candidate lensed pairs involving \ac{BNS} or \ac{NSBH} events are not analyzed and ranked.

We additionally employ a \ac{ML}-based binary classification scheme to rapidly provide a probability of class membership (lensed or unlensed) for a given candidate \ac{BBH} pair \citep{Goyal:2021hxv}.
Such an analysis not only serves as an independent method to rank candidate pairs but also provides a quantitative significance to pairs for which source-parameter inference samples are unavailable.

Q-transform-based \citep{Chatterji:2004qg} time--frequency maps of strongly lensed \acp{BBH} are expected to have similar shapes, although the signal energy in each time-frequency tile will differ between images. 
Furthermore, as mentioned earlier, their sky localisations will overlap.
Exploiting these facts, \ac{ML} models that take Q-transforms and \texttt{Bayestar} \citep{Singer:2015ema} sky localisations as inputs, are built.
These models use a \texttt{DenseNet} \citep{Huang:2016ML} architecture \citep[with several layers pre-trained on the \texttt{ImageNet} dataset;][]{Deng:2009ML}, and \texttt{XGBoost} \citep{Chen:2016btl} algorithms, trained on lensed and unlensed \ac{BBH} signals injected in Gaussian noise (for details on the \ac{ML} training set, see \citet{Goyal:2021hxv}) 
The outputs of the individual models are then combined to provide a probability that a candidate pair is lensed or unlensed. 

To convert this probability to an \ac{FPP}, we construct a background distribution of \ac{ML} probabilities using a population of unlensed \ac{BBH} events injected in Gaussian noise characterized by the O3a representative \ac{PSD} -- the same as was used for the posterior overlap statistic.  
This \ac{PSD} is found to be sufficiently similar to the averaged O3 \ac{PSD} for the estimation of the background distribution so as not to change the preliminary selection of candidate pairs. 
The \ac{BBH} population is identical to the one used by the posterior overlap statistic analysis to construct its corresponding background distribution. 
Furthermore, the sky localisations used to rank candidate pairs come from the same PE analysis used to estimate the posterior overlap statistic \footnote{Note that \texttt{Bayestar}, which is used to assign \ac{ML} probabilities to real-event candidate pairs, is expected to provide sky localisations that are similar to those provided by this \ac{PE} analysis.}.   

\begin{figure}[h!]
	\centering
	\includegraphics[width=\columnwidth]{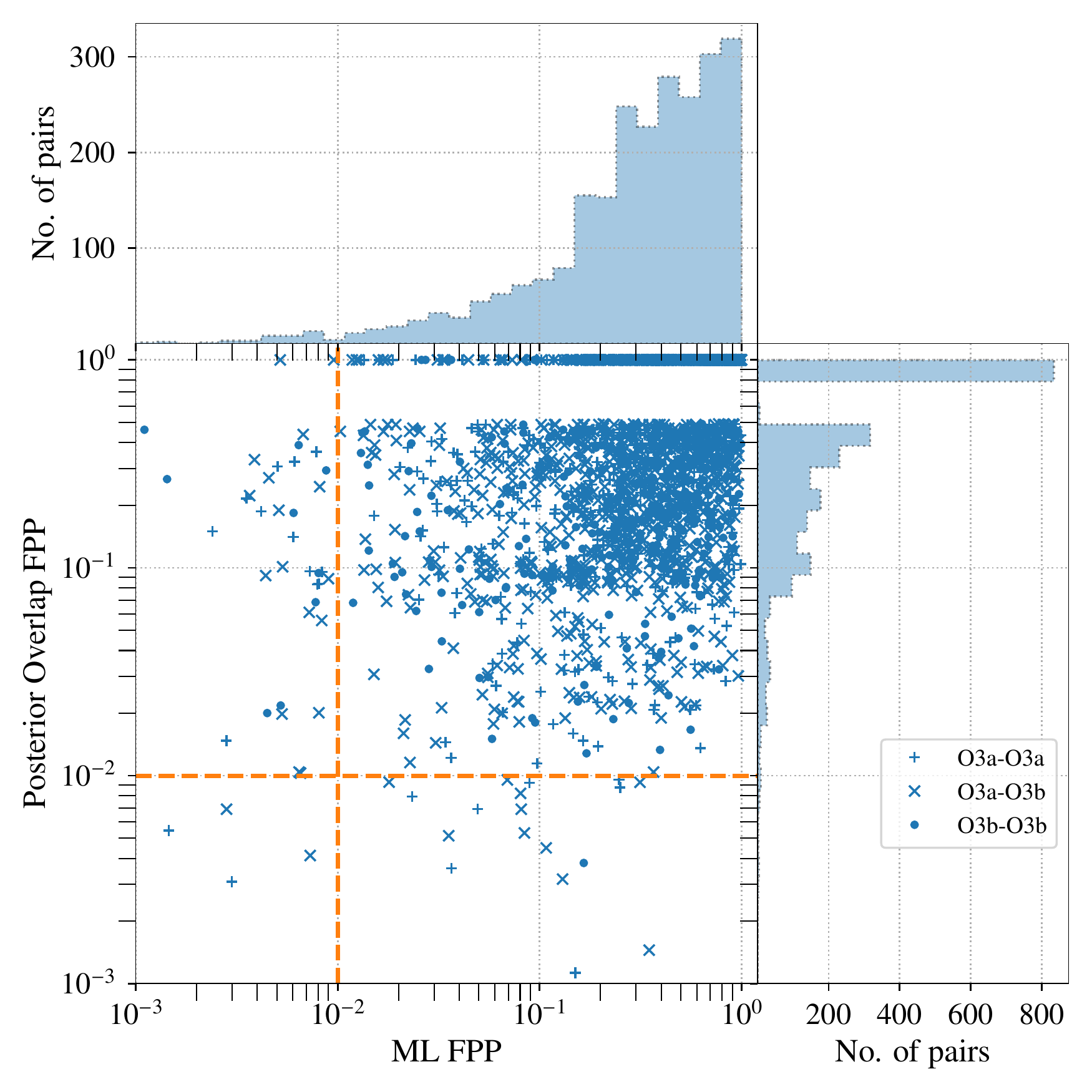}
	\caption{
		The \acp{FPP} of each lensed candidate pair constructed from the set of \ac{GW} events that exceed an astrophysical probability \citep{Farr:2013yna, Kapadia:2019uut} threshold of $0.5$, as evaluated using the $\BLU~$ and \ac{ML} classification statistics.
		Orange dashed lines that correspond to an \ac{FPP} threshold of $10^{-2}$, are also placed.
		Pairs whose $\BLU$-based or \ac{ML}-based \acp{FPP} fall below this threshold are selected for additional joint parameter estimation analyses.
		$\BLU < 10^{-6}$ has been mapped to an \ac{FPP} of 1, which is reflected in the gap along the vertical axis between $0.4$ and $1$.
	}
	\label{fig:POML}
\end{figure}

A plot comparing the \acp{FPP} assigned by the posterior overlap and \ac{ML} analyses is shown in Fig.~\ref{fig:POML}.  Candidates that have either a posterior-overlap-assigned \ac{FPP} or \ac{ML}-assigned-\ac{FPP},  (or both),  that are smaller than $1\%$, are selected for more comprehensive Bayesian analyses.

    \subsection{Joint Parameter Estimation}
    \label{sec:jointpe}
    Similar to the analysis of O3a data~\citep{LIGOScientific:2021izm}, we perform a joint PE analysis for the most relevant candidate lensing pairs.
We follow up on the pairs that display low FPP in their posterior overlap or \ac{ML} classification scheme. 
These are pairs within the whole of O3, but we only consider here those with at least one event in O3b since pairs in O3a were studied in \citet{LIGOScientific:2021izm}.
We use two complementary pipelines: \texttt{golum} \citep{Janquart:2021qov} and \texttt{hanabi} \citep{Lo:2021nae}. 
Both pipelines use the nested sampling algorithm \texttt{dynesty}~\citep{10.1093/mnras/staa278}, and implement the joint PE with the help of \texttt{bilby}~\citep{Ashton:2018jfp, Romero-Shaw:2020owr}. 

\texttt{golum} \citep{Janquart:2021qov} is a joint PE tool where the workload is reduced by analyzing the two images, under the lensed hypothesis, in two successive stages.
The first image is characterized by the same parameters of the unlensed case (where the time of coalescence and the luminosity distance are the observed ones) with an additional Morse factor. 
The second image is then analyzed using (samples of) the posterior from the first image as the prior and linking the parameters modified by lensing through three lensing parameters: a time difference, a relative magnification, and a Morse factor difference. 
The final coherence ratio $\Clu$ is the ratio of the product of the evidences for the two runs under the lensed hypothesis and the product of evidences for the two images analyzed under the unlensed hypothesis. 

\texttt{hanabi}~\citep{Lo:2021nae} first performs a joint inference on a signal pair by constructing a joint likelihood function that is a product of the likelihood function for each individual event, with a joint prior distribution. 
The latter is defined for a set of joint parameters that can simultaneously describe both signals if they are truly lensed, for example, the masses and the spins, as well as a set of parameters that are different for each of the signals such as the time of arrival, the apparent luminosity distance, and the Morse phase factor associated to each of the lensed signals. 
The joint parameter space is explored with the package \texttt{hanabi.inference}~\citep{Lo:2021nae}.
The inference result is then reweighted with an astrophysically motivated prior distribution; for example, the astrophysical prior distribution for the redshifted component masses would be dependent on both the population model for the intrinsic BBH masses and the redshift distribution of the sources. 
However, the true source redshift cannot be determined from \ac{GW} observations alone since the true source redshift is degenerate with the magnification from strong lensing. 
To compute the Bayes factor $\Blu$, the source redshift, which serves as a hyper-parameter for the signal pair, must be marginalized over. 
Selection effects enter as a normalization constant to the marginal data likelihood.
This procedure is implemented in \texttt{hanabi.hierarchical} with the help of \textsc{gwpopulation} \citep{Talbot:2019okv}.
The ratio of unnormalized evidences calculated under the lensed hypothesis and the unlensed hypothesis using this astrophysical prior is referred to as the population-weighted coherence ratio $\left.\cohratio\right|_\mathrm{pop}$, while the ratio of normalized evidences that accounts for both population prior and selection effects is referred to as the Bayes factor $\Blu$ in this analysis. 
We follow our fiducial \ac{SIS} lensing model when computing the magnification prior~\citep{LIGOScientific:2021izm}. 
This analysis however does not impose any informative prior on the time delay or the image types from the lensing model.

Both pipelines use \IMRXPHM~\citep{Pratten:2020ceb} as the waveform model, with an additional Morse phase applied to each of the waveform polarizations in the frequency domain. 
Other inputs, such as the power spectral density estimates and the calibration envelopes, are chosen to match the analyses done in the GWTC-3 catalog paper \citep{LIGOScientific:2021djp}.
Following the same prescriptions of the other analyses, we fix the BBH population model to the Power-Law + Peak model for the primary masses and the merger rate history to Madau--Dickinson star-formation rate \citep{Madau:2014bja} normalized by the median GWTC-3 rate \citep{LIGOScientific:2021psn}.

\begin{figure*}[ht]
	\centering
	\includegraphics[width=2\columnwidth]{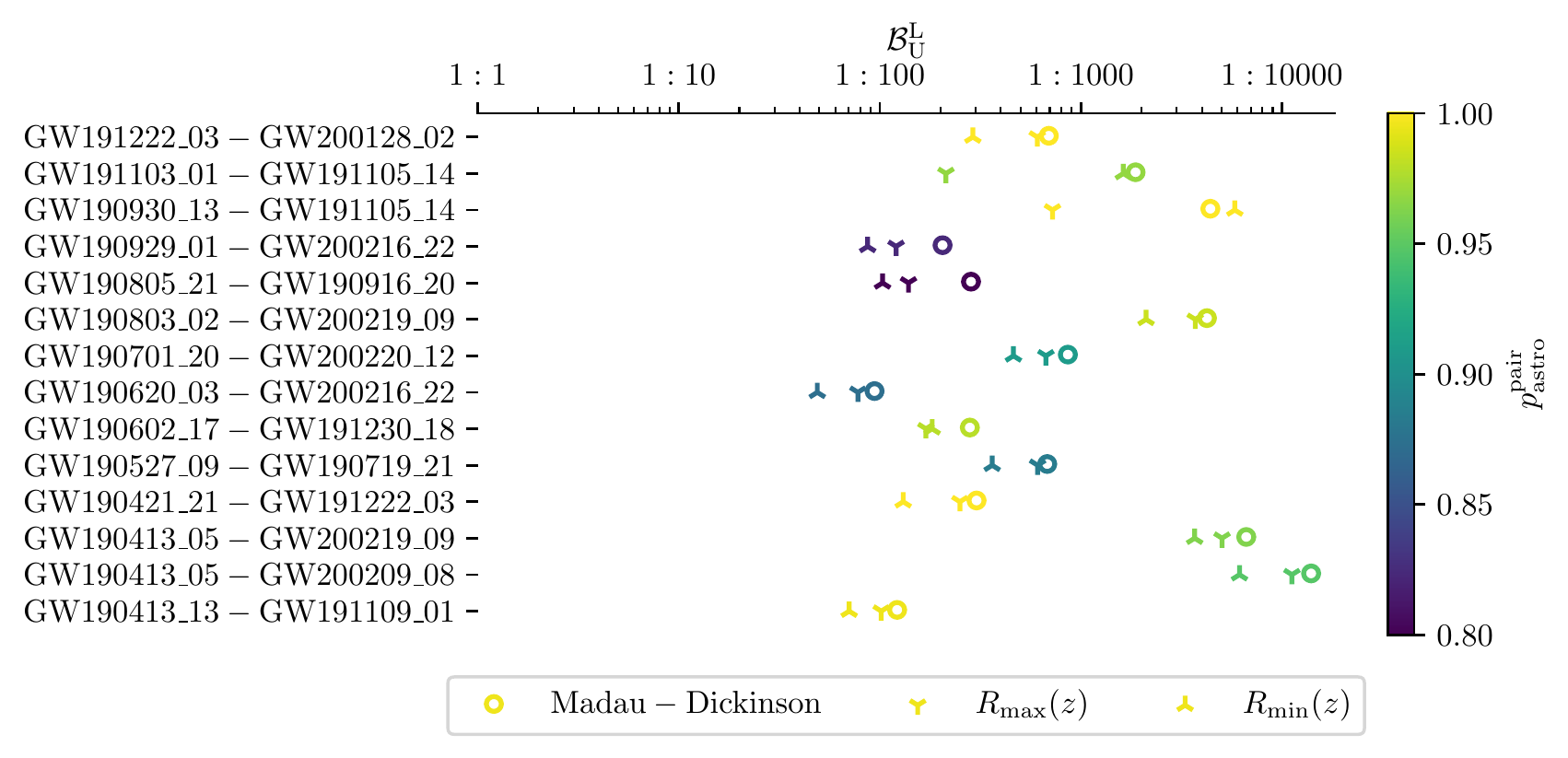}
	\caption{
		Bayes factors $\Blu$ from \texttt{hanabi} for the highest-ranked multiple-image candidate pairs.
		As a check on the robustness of our results, we show the Bayes factors calculated using three different merger rate density models, namely the fiducial model tracking the Madau--Dickinson star-formation rate \citep{Madau:2014bja}, and also the $R_{\rm min}(z)$ and $R_{\rm max}(z)$ model introduced in \citet{LIGOScientific:2021izm}.
		The color for each marker represents the value of $p_{\rm astro}^{\rm pair}$ for each pair, which is the probability that both of the signals from a pair are of astrophysical origins and not from terrestrial sources.
	}
	\label{fig:bayes_factor_from_hanabi}
\end{figure*}

Taking advantage of \texttt{golum}'s rapid joint PE, we analyze the 75 pairs of candidates highlighted by posterior overlap and \ac{ML}.
For each of them, we compute the coherence ratio, which accounts for the probability ratio of the lensed and unlensed hypotheses without including selection effects and population priors. 
We find that there is a wide range of $\log_{10}(\Clu)$ values, with a peak slightly above zero. 
This comes from the fact that this analysis considers only triggers already flagged by the posterior overlap and \ac{ML} analyses. 
As a consequence, the analysis is biased towards the higher values.
Nevertheless, a significant proportion of events flagged with the \ac{ML} pipeline and the posterior overlap pipeline are disfavored, having $\log_{10}(\Clu)<0$. 
When comparing the highest coherence ratio found in the data, $\log_{10}(\Clu) = 2.5$, with a background of unlensed events, we find that it is well within the expected values, with $1\%$ of the background events having larger $\Clu$. 
This background is computed for a population of compact binaries that follows the mass, spin and redshift distribution of GWTC-3 \citep{LIGOScientific:2021psn}.  
This large number of positive  $\log_{10}(\Clu)$ is consistent with the high number of expected false alarms~\citep{Wierda:2021upe,Caliskan:2022wbh}. 
For those pairs with the highest coherence ratio, we follow up with the \texttt{hanabi} pipeline for a total of 17 pairs. 
Our main results are presented in Fig. \ref{fig:bayes_factor_from_hanabi}, where the left column indicates the event pairs and the horizontal axis their $\Blu$.
There we can observe that none of the event pairs shows support for the lensing hypothesis, i.e. all $\Blu<1$. 
The pair with highest $\Blu$ is GW190620\_030421 -- GW200216\_220804, for an evidence against lensing of $\sim 1/100$ with the fiducial merger rate density model following the Madau-Dickinson star-formation rate.
As a robustness check of how using different merger rate density models would change the results, we repeat the calculations using two more models, namely $R_{\rm min}(z)$ and $R_{\rm max}(z)$ from our previous O3a analysis \citep{LIGOScientific:2021izm} that minimally and maximally bracket many existing population-synthesis results~\citep{Belczynski:2005mr,Belczynski:2010tb,Dominik:2013tma,Eldridge:2018nop}.
We see that while the exact values for the Bayes factor change with the use of different merger rate density models, the conclusion remains that there is no support for the lensing hypothesis in any of the event pairs analyzed.
To further assess the significance of these pairs we also include a color code to indicate the probability of having an astrophysical origin $p_\mathrm{astro}^\mathrm{pair}$, defined as the product of the highest $p_\mathrm{astro}$ of each event reported in the GWTC-3 catalog paper \citep{LIGOScientific:2021djp} by different pipelines.
In conclusion, we find no evidence of multiply imaged events.

    \subsection{Type II image search}
    \label{sec:typeII}
    In addition to the search for strong-lensing identifying multiple images, we also look for the distortions that lensing introduces in type II images \citep{Ezquiaga:2020gdt}.
This is because the frequency-independent phase shift that each image acquires becomes a frequency-dependent time delay for different frequency components. 
Therefore, for signals containing different measurable spherical harmonic modes, as recently detected in GW190412~\citep{LIGOScientific:2020stg}, GW190814~\citep{GW190814}, and other events~\citep{LIGOScientific:2021djp}, the overall lensed waveform can be distorted. 
The extent of the distortion is subject to the power in modes beyond the quadrupole radiation. 
As a consequence, we do not expect to see these distortions in the majority of the lensed events with current sensitivities. 
However, if not searched for, they might be mistaken with deviations from general relativity~\citep{Ezquiaga:2022nak}.

\begin{figure}[h!]
	\centering
	\includegraphics[width=\columnwidth]{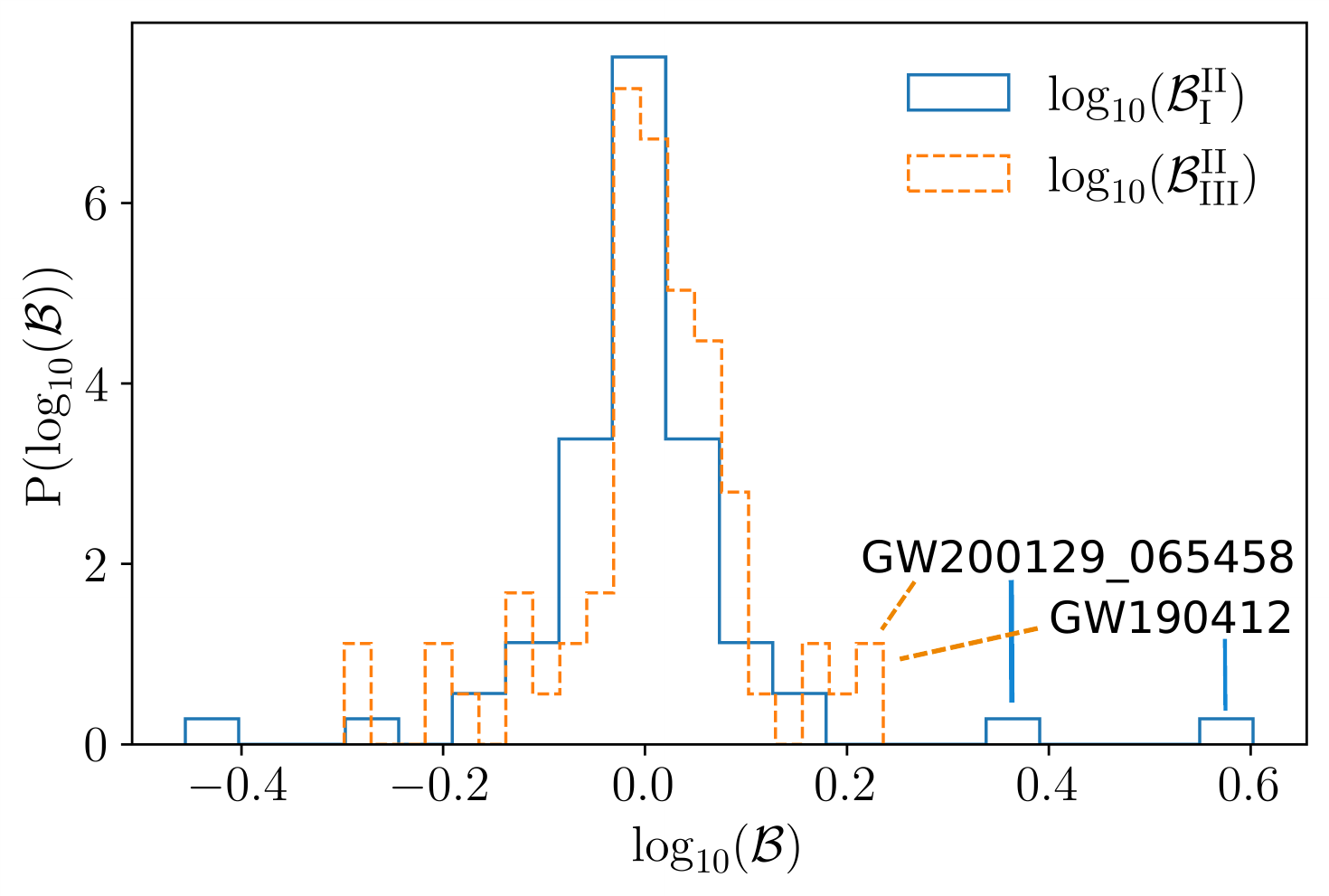}
	\caption{
		Distribution of Bayes factors comparing different image type hypotheses for the 10 most relevant events.
		We compare the probability of being type II vs. type I (blue-solid histogram) and of being type II vs. type III (orange-dashed histogram).
		Only type II images display waveform distortions and for that reason, we do not compare type III vs. type I.
	}
	\label{fig:MorsePhase_Blu}
\end{figure}

To look for these distortions, we use \golum \citep{Janquart:2021qov}. 
Within GWTC-3 we identify 10 events whose posterior has some information about the Morse phase, either by favoring or disfavoring the distortions of the type II image by more than $4\%$ with respect to normality, i.e. the probability of each image type $p(n_j)$ is $p(n_j)>0.37$ or $p(n_j)<0.29$. 
We summarize the evidence of one image type versus another in Fig.~\ref{fig:MorsePhase_Blu}. 
Since only type II images display waveform distortions, we only compute the Bayes factors of the type-II-vs-I and the type-II-vs-III hypotheses.
As can be seen in Fig.~\ref{fig:MorsePhase_Blu}, only a few events display a preference for one image type versus the other one.
This is expected given the \ac{SNR} of these events and their power in higher multipole moments. 
However, GW190412 and GW200129\_065458 present higher evidence for type II images. 
For GW190412 we find a $\log_{10}$ Bayes factor for type II vs. I of $0.60 \pm 0.16$ and for type II vs. III of $0.22 \pm 0.16$. 
For GW200129\_065458 we find $0.38\pm0.14$ and $0.24\pm0.14$ for type II vs. I and type II vs. III respectively. 
These events have possible super-threshold counterparts but those were discarded by the \golum analysis. 
In addition, we have also searched for sub-threshold triggers associated with these events, but found none.

To assess the significance of the type II images, we follow up on GW190412 and GW200129\_065458 performing a simulation campaign of type I and type II images. 
GW190412 simulations show that indeed this event has enough power in higher multipole moments to favor the type II hypothesis so that it could meaningfully test that hypothesis and would favor it if it were true. 
For GW200129\_065458, however, that is not the case. 
Moreover, GW200129\_065458 might have a significant glitch under subtraction~\citep{Payne:2022spz}.
The preference of GW190412 for a type II image could be just a systematic effect due to the waveform modeling, especially since this event falls in challenging parts of the parameter space~\citep{LIGOScientific:2020stg,Colleoni:2020tgc,Hannam:2021pit}. 
For this reason, we repeat the analysis with different waveform families from our fiducial \IMRXPHM model~\citep{Pratten:2020ceb}. 
We find that the preference for a type II image remains when using \textsc{SEOBNRv4PHM}~\citep{Ossokine:2020kjp} or \textsc{IMRPhenomPv3HM}~\citep{Khan:2019kot}. 
The same conclusion holds when using different noise realizations for the simulations. 
Details on these simulation campaigns can be found in Appendix \ref{sec:app_typeII}.

Although we find a mild preference for the type II image hypothesis in GW190412, we find that this analysis cannot provide conclusive evidence of strong lensing.
However, our techniques and pipeline will be relevant for future observing runs when high-\ac{SNR} events display stronger evidence of higher-order modes.

\section{Microlensing Effects}
\label{sec:microlensing}
When the characteristic wavelengths of \acp{GW} are comparable to the Schwarzschild radius of a lens ($\lambda_\mathrm{GW}\sim R^\mathrm{lens}_\mathrm{Sch}$), we may observe frequency-dependent magnification of the waveform that can inform us about the lens model~\citep{Takahashi:2003ix,Cao:2014oaa,Jung:2017flg,Lai:2018rto,Christian:2018vsi,Dai:2018enj,Diego:2019lcd,Diego:2019rzc,Pagano:2020rwj,Cheung:2020okf,Cremonese:2021puh,Caliskan:2022hbu}.
Since the \acp{GW} of sources such as \acp{BBH} sweeps through a wide range of frequencies, these beating patterns can reveal the presence of intervening microlenses.
In the sensitive range of ground-based detectors, these effects are expected for objects up to $\sim 10^5 M_{\sun}$, which includes stellar-mass objects and intermediate-mass \acp{BH}.

Objects that can cause these microlensing effects are predominantly found in larger structures.
Therefore we expect that realistic microlensing due to a field of microlenses embedded in an external macromodel potential such as galaxies and galaxy clusters causes complex effects on the unlensed waveforms~\citep{Diego:2019lcd}.
While the effects of these systems on \ac{GW} signals have been studied \citep{Diego:2019rzc,Cheung:2020okf,Mishra:2021xzz,Yeung:2021roe}, the resulting waveforms are computationally costly to evaluate.
Nevertheless, in the absence of specific knowledge of the matter distribution along the travel path and to keep the problem computationally tractable, we assume that the beating patterns are caused by isolated point masses as a first approximation.
In this case, the microlensed waveform $h^\mathrm{Micro}$ can be related to the unlensed waveform $h^\mathrm{U}$ according to
\begin{equation}
	h^\mathrm{\rm Micro}(f;\theta, \MzL, y) = h^\mathrm{U}(f;\theta) \, F(f; \MzL, y)\,,
	\label{eq:wl_lens_hyp}
\end{equation}
where $\theta$ represents the set of parameters defining an unlensed \ac{GW} signal, \mbox{$\MzL = M_\mathrm{L} (1+\zL)$} is the redshifted lens mass, $y$ is the dimensionless impact parameter, and $F(f; \MzL, y)$ is the frequency-dependent lensing magnification factor \citep[e.g.,][]{Takahashi:2003ix}.

Similar to \citet{LIGOScientific:2021izm}, we perform Bayesian inference on all events from O3b using the unlensed signal model $h^\mathrm{U}$ and the microlensing signal model $h^\mathrm{Micro}$.
In particular, we use \texttt{bilby}~\citep{Ashton:2018jfp, Romero-Shaw:2020owr} and the nested sampling algorithm \texttt{dynesty}~\citep{10.1093/mnras/staa278}.
Data products such as strain data and \acp{PSD} are the same as for GWTC-3 and between the two signal models \citep{LIGOScientific:2021djp}
For the \ac{GW} parameters, we use the same priors as GWTC-3, while the prior on the lens mass $\MzL$ is log uniform in the range $[1$--$10^5~\Msun]$ and the prior on the impact parameter is $p(y)\propto y$ between $[0.1,3]$.
All events were analyzed using \IMRXPHM~\citep{Pratten:2020ceb}.

The process yields posterior probability distributions of $\theta$ or $\left\{ \theta, \MzL, y \right\}$ for the unlensed and lensed signal models, respectively.
Moreover, we compute the evidence ratio between the microlensed and unlensed signal models, better known as the Bayes factor $\BMLU$.
\begin{figure}[h!]
	\centering
	\includegraphics[width=\columnwidth]{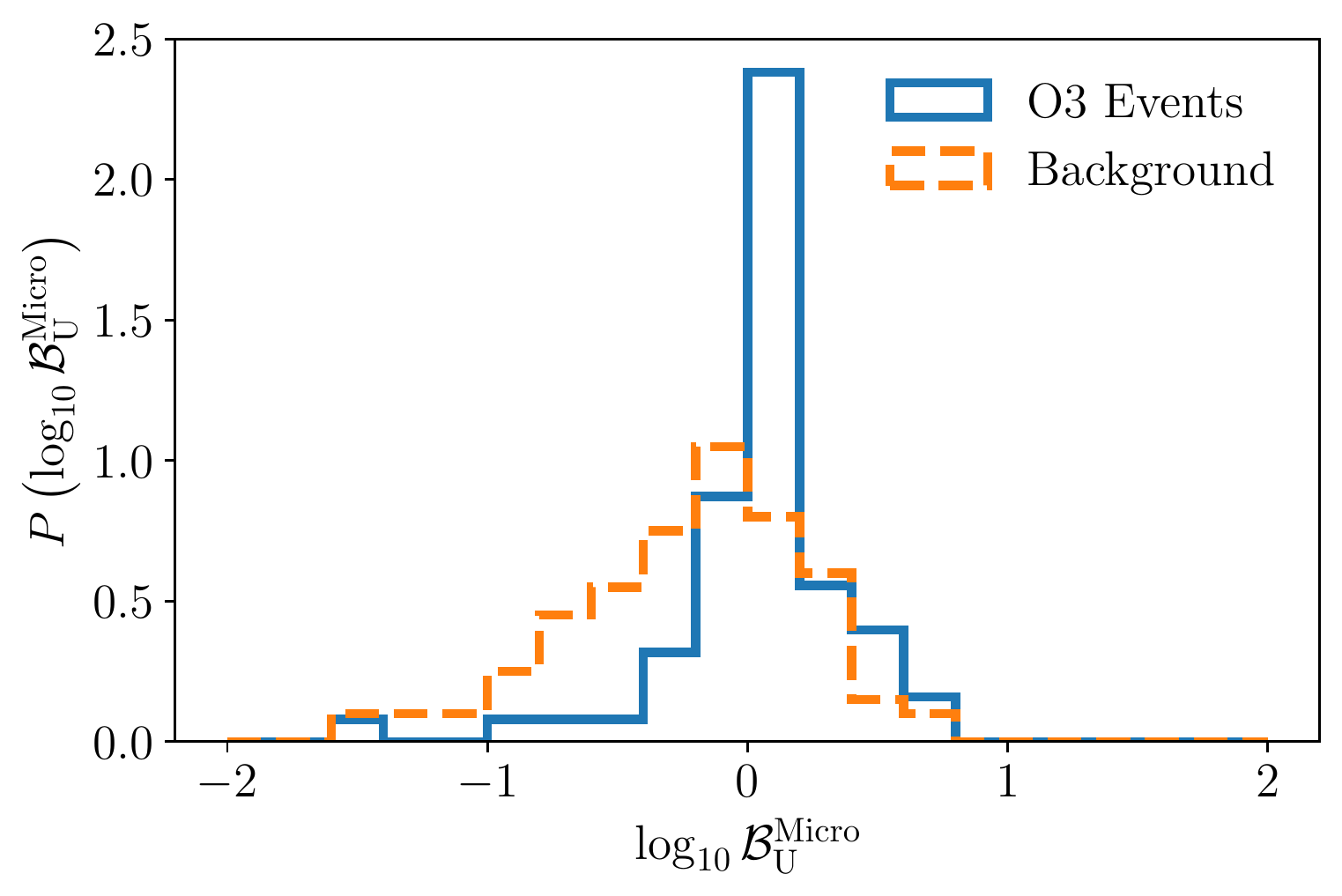}
	\caption{Distribution of microlensing $\log_{10}$~Bayes factors $\BMLU$ for all events in O3 (blue, solid line) and simulated unlensed signals (orange, dashed line) from \citet{LIGOScientific:2021izm}.}
	\label{fig:micro_BLUdistr}
\end{figure}
Fig.~\ref{fig:micro_BLUdistr} shows the distribution of $\log_{10} \BMLU$ for all the events in O3 and simulated unlensed signals from \citet{LIGOScientific:2021izm}.
The distribution of $\log_{10} \BMLU$ is primarily clustered around 0 and the distribution for O3 events does not extend to significantly higher values than the distribution for simulated signals.
The marginalized posteriors of the microlensing parameters are shown in Appendix \ref{sec:app_micro}.
We conclude that there is no compelling evidence for the presence of microlensing signatures.

\section{Implications}
\label{sec:implications}
In this section, we consider some of the implications that derive from the search for lensing signatures.
We first forecast the number of detectable strongly lensed events based on the latest knowledge on the merger-rate density (Sec.~\ref{sub:lensing_rate_strong}).
Next, we infer upper limits on the strong lensing rate using the non-detection of resolvable strongly lensed \ac{BBH} events (Sec.~\ref{ssec:implications_of_non_detection}).
Finally, we use the non-detection of microlensing to infer the compact dark matter fraction in the Universe (Sec.~\ref{sub:dm-fraction}).

\subsection{Strong lensing rate}
\label{sub:lensing_rate_strong}
We predict the rate of lensing using the standard methods outlined in the literature~\citep{Ng:2017yiu,Li:2018prc,Oguri:2018muv,Xu:2021bfn,Mukherjee:2021qam,Wierda:2021upe}, at galaxy and galaxy-cluster lens mass scales.
To model the lens population, we need to choose a density profile and a mass function. 
We adopt the \ac{SIS} density profile for both galaxies and galaxy clusters.
Moreover, we use the velocity dispersion function from the Sloan Digital Sky Survey \citep{Choi:2007a} for galaxies and the halo mass function from \citet{Tinker:2008ff} for clusters which have also been used in other lensing studies \citep[e.g.,][]{Oguri:2010a,Robertson:2020mfh}.
The \ac{SIS} profile can accurately describe lensing by galaxies, but the mass distribution of clusters tends to be more complicated.
Nevertheless, \citet{Robertson:2020mfh} have demonstrated that the \ac{SIS} model can reproduce the lensing rate predictions from a study of numerically simulated cluster lenses.
Thus, we adopt the same model for both galaxies and galaxy clusters.

Under the \ac{SIS} model, we obtain two images with different magnifications and arrival times.
The rate of strong lensing is given by
\begin{equation}
	\begin{split}
		\mathcal{R}_{\rm lens}= \int
		& \frac{\mathrm{d}N(M_{\mathrm{h}},\zL)}{\mathrm{d}M_{\mathrm{h}}} \frac{\mathrm{d}D_{\mathrm{c}}} {\mathrm{d}\zL}
		\frac{\mathcal{R}_{\mathrm{m}}(z_{\mathrm{m}})}{1+z_{\mathrm{m}}} \frac{\mathrm{d}V_{\mathrm{c}}}{\mathrm{d}z_{\mathrm{m}}}\, \sigma(M_{\mathrm{h}},\zL,z_{\mathrm{m}},\rho, \rho_{\mathrm{c}}) \\
		&\times p(\rho|z_{\mathrm{m}}) ~\mathrm{d}\rho~\mathrm{d}z_{\mathrm{m}}~\mathrm{d}\zL~\mathrm{d}M_{\mathrm{h}} \,, \label{eq:rates1}
	\end{split}
\end{equation}
where $\mathrm{d} N(M_\mathrm{h},\zL) / \mathrm{d}M_{\mathrm{h}}$ is the differential comoving number density of lensing halos in a halo mass shell at lens redshift $\zL$, $D_{\mathrm{c}}$ and $V_\mathrm{c}$ are the comoving distance and volume, respectively, at a given redshift, $\mathcal{R}_\mathrm{m}(z_\mathrm{m})$ is the total comoving merger rate density at redshift $z_\mathrm{m}$, (1+$z_\mathrm{m}$) accounts for the cosmological time dilation, $p(\rho\mid z_{\rm m})$ is the distribution of \ac{SNR} at a given redshift, $\rho_\mathrm{c}$ is the network \ac{SNR} threshold, and $\sigma$ is the lensing cross-section which indicates, as a function of its various arguments, how efficiently strong lensing will occur. 
We model the mass distribution of \acp{BBH} following the results for the \textsc{Power Law + Peak} model of~\citet{LIGOScientific:2021psn}.
We consider a merger rate density model that assumes the Madau--Dickinson ansatz~\citep{Madau:2014bja} that is consistent with recent results from GWTC-3.
Moreover, we make use of the absence of a detected \ac{SGWB} to further constrain the merger rate density \citep{LIGOScientific:2021psn}.
For consistency with previous analyses \citep[e.g.,][]{Abbott:2021xxi}, we take the Hubble constant from Planck 2015 observations to be $H_0=67.9\ {\rm km\ s^{-1}\ Mpc^{-1}}$ \citep{Planck_2015}.
Furthermore, we choose $\rho_\mathrm{c}=8$ as a point estimator of the detectability of \ac{GW} signals. 
We find this choice to be consistent with the search results in~\citet{GWTC2} and Sec.~\ref{sec:subthreshold}, and we estimate its impact to be subdominant compared to other sources of uncertainty.

\begin{table*} 
	\caption{
		\label{tab:lensedrates0_min}
		Expected fractional rates of observable lensed double events at current LIGO--Virgo sensitivity.
	}
	\begin{center}
		\begin{tabular}{lcccc}
			\hline\hline
			Merger Rate Density  & \multicolumn{2}{c} {Galaxies}  & \multicolumn{2}{c}  {Galaxy Clusters}\\ 
			Model &  $R_{\rm D}$ & $R_{\rm S}$ & $R_{\rm D}$ & $R_{\rm S}$ \\ 
			\hline\hline
			GWTC-3+Stochastic  & \RateGalaxyDouble & \RateGalaxySingle & \RateClusterDouble & \RateClusterSingle \\
			\hline
		\end{tabular}
	\end{center}
	\tablecomments{
		This table lists the relative rates of lensed double events expected to be observed by LIGO--Virgo at the current sensitivity where both of the lensed events are detected ($R_{\rm D}$) and only one of the lensed events is detected ($R_{\rm S}$) above the \ac{SNR} threshold.
	The rates encompass a 90~percent credible interval.
	We show the rate of lensing by galaxies ($\sigma_{\rm vd}=10$--$300 \,{\rm km~s}^{-1}$) and galaxy clusters ($\log_{10}(M_{\rm halo}/M_\odot) \sim 14$--$16$) separately. 
}
\end{table*}

In Table~\ref{tab:lensedrates0_min}, we show our estimates for the relative rate of lensing expected to be observed by the LIGO--Virgo network of detectors.
The results are shown separately for galaxy-scale and cluster-scale lenses.  
Furthermore, these rates are calculated for events that are doubly lensed and for two cases:
when only a single event (i.e., the brighter one) is detected (S), and when both of the doubly lensed events are detected (D).
The expected fractional rate of lensing (i.e. the lensed to unlensed rate) spans the range $\mathcal{O}(10^{-4}$--$10^{-3})$, depending on the merger rate density assumed.
We estimate the fractional rate of observed double (single) events for galaxy-scale lenses to lie in the range \RateGalaxyDouble~(\RateGalaxySingle).
Similarly, for cluster-scale lenses, the fractional rate is estimated to be in the range of \RateClusterDouble~(\RateClusterSingle), typically lower than the rates on galaxy scales. 
These estimates suggest that observing a lensed double image is unlikely at the current sensitivity of the LIGO--Virgo network of detectors. 
Nevertheless, at design sensitivity and with future upgrades, standard forecasts suggest that the possibility of observing such events might become significant~\citep{Ng:2017yiu,Li:2018prc,Oguri:2018muv,Xu:2021bfn,Mukherjee:2021qam,Wierda:2021upe}.
Compared with other lens models, our lensing rates are consistent with those predicted for \ac{SIE} models \citep[e.g.,][]{Oguri:2018muv,Xu:2021bfn,Wierda:2021upe}.

\subsection{Implications from the non-observation of strongly lensed events} 
\label{ssec:implications_of_non_detection}
The absence of any detections of strongly lensed \ac{GW} events before and during O3 provides a complementary way to constrain the merger rates of compact objects at high redshift.
The detection of individual \ac{GW} events has enabled measurement of the low redshift ($z<1$) merger rate \citep{LIGOScientific:2021psn}.
However, the high redshift merger rate of \ac{GW} sources is not yet measured directly, and we have only been able to place an upper limit on it from the absence of a detection of the \ac{SGWB} \citep{KAGRA:2021kbb}.
The absence of such a detection naturally leads to a bound on the lensing rate expected from GWTC-3 \citep{Mukherjee:2020tvr, Buscicchio:2020cij}.

\begin{figure}
	\centering
	\includegraphics[width=\columnwidth]{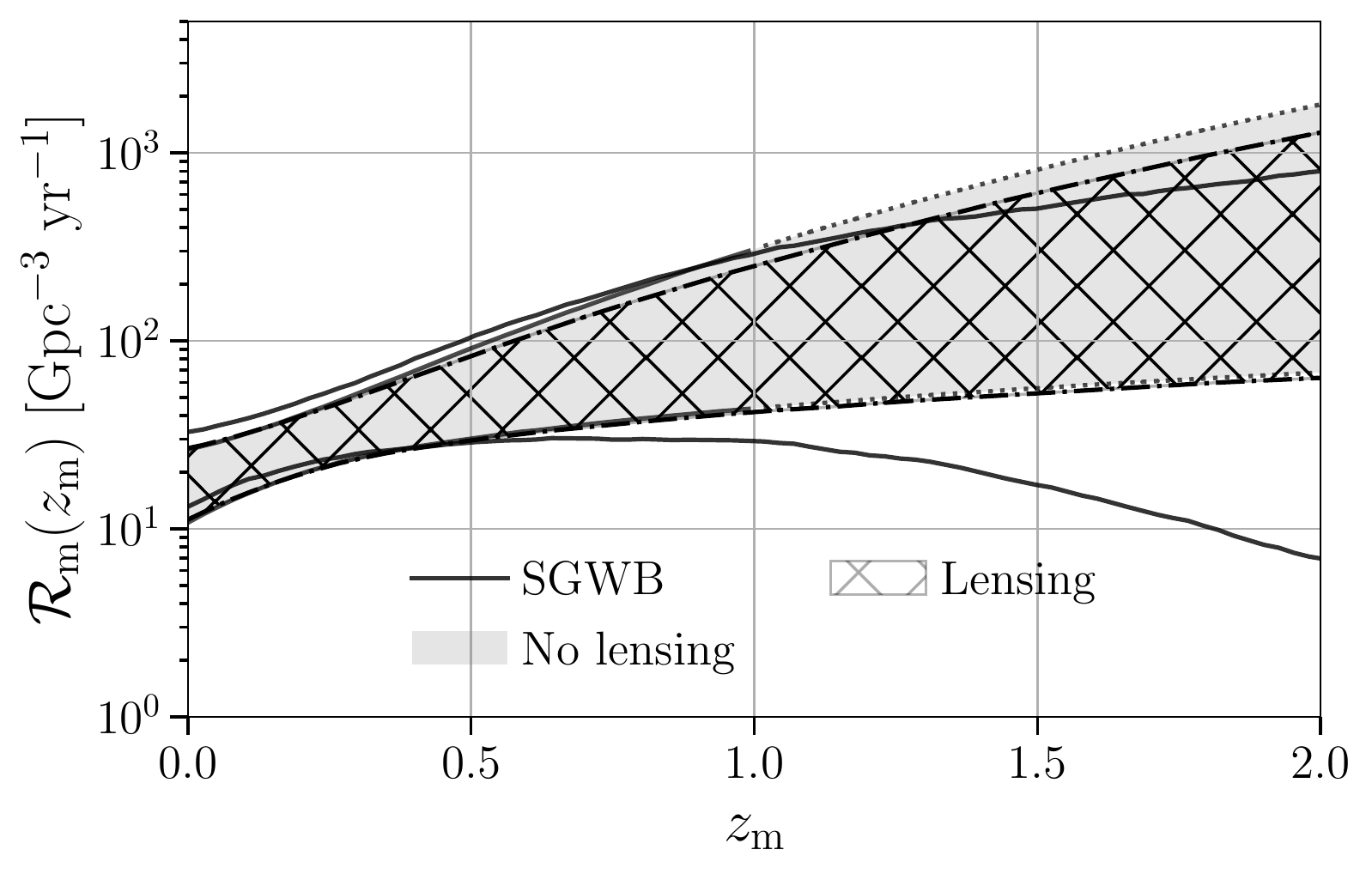}
	\caption{
		Merger rate density as a function of redshift based on the GWTC-3 results without lensing constraints (grey) and with lensing constraints (cross-hatching) included. 
		For clarity, we show only the results for galaxy-scale lenses.
		Because lensed detections may occur at higher redshifts than unlensed events, their non-observation can be used to constrain the rate of mergers at higher redshifts.
		The `No lensing' results shown here do not include constraints derived from the absence of an \ac{SGWB}. The latter constraints are shown separately by the solid black curves.
	}
	\label{fig:merger_rate_constraints_from_strong_lensing}
\end{figure}

By using the same power-law form for the merger rate as that used in Sec. \ref{sub:lensing_rate_strong}, but extended up to $z=2$, we obtain limits on the merger rate at redshift $z>1$ from the absence of detections of strongly lensed events.
The corresponding constraints ($90 \%$ credible intervals) are shown in Fig. \ref{fig:merger_rate_constraints_from_strong_lensing} as the cross-hatched region bounded by the dash-dotted curves.
The changes in the upper bound of the merger rates are driven by the absence of detected lensing events, whereas the lower bound is driven by the low-redshift constraints on the merger rate.
For comparison, the current limits on the merger rate from GWTC-3 up to redshift $z=1$ \citep{LIGOScientific:2021psn}, with the bounding curves extrapolated to higher redshifts $z>1$, are shown as the grey shaded region bounded by the dotted curves.
For further comparison, we have also plotted the solid black curves which show the current constraints from the absence of detection of the \ac{SGWB} \citep{KAGRA:2021kbb}.
The upper bounds on the merger rate from lensing are more stringent than the bounds from GWTC-3 at high redshift \citep{LIGOScientific:2021psn}, and are also comparable with the bounds from the \ac{SGWB} for redshifts $z< 1.2$.
The slight difference between the constraints on the merger rates at low redshift derived from the \ac{SGWB} \citep{KAGRA:2021kbb} and from GWTC-3 \citep{LIGOScientific:2021psn} arise because the bounds from the \ac{SGWB} are obtained here using the previous constraints on the merger rate at low redshift derived using GWTC-2 \citep{Abbott:2020gyp}.

\subsection{Constraints on compact dark matter from gravitational-wave microlensing}
\label{sub:dm-fraction}
Objects whose size is comparable to their gravitational radius, and that cause microlensing effects on \ac{GW} signals, could be candidates for dark matter. 
Although their abundance is heavily constrained by several astronomical observations~\citep{Carr:2020xqk, Carr:2020gox}, the possibility of their contributing to dark matter cannot be ruled out in several mass windows.

Here we use the non-observation of microlensing effects on the \ac{GW} signals detected by LIGO and Virgo to constrain the fraction of dark matter contributed by compact objects in the mass range $\sim 10^2$--$10^5~{M_\odot}$~\citep{Jung:2017flg,Urrutia:2021qak,Basak:2021ten}.
The essential idea is that if a significant fraction of dark matter is in the form of compact objects, they would introduce detectable microlensing signatures on the \ac{GW} signals that we observe. 

Assuming that lensed and unlensed events occur as Poisson processes, we compute the posterior distribution on the lensing fraction ($u \equiv \LambdaL/\Lambda$), defined as the ratio of Poisson means of lensed events to the total number of detected events. This is then used to compute the posterior of the fraction of compact dark matter ($\fdm \equiv \Omega_\mathrm{CO}/\Omega_\mathrm{DM}$) \citep{Basak:2021ten}.
We take that a total of $N = 67$ \ac{BBH} mergers are detected during the O3 run \footnote{These are the events cataloged in GWTC-3 that do not contain a neutron star component.}, and none of them is lensed (i.e., $\NL = 0$). 
We then estimate the posterior distribution of the lensing fraction $u$.
Finally, the posterior of $\fdm$ can be computed as 
\begin{equation}
p(\fdm \mid \{\NL = 0, N\})  =  p(u \mid \{\NL = 0, N\}) \left| \frac{\mathrm{d} u}{\mathrm{d}\fdm} \right|,
\label{eq:fdm_posterior}
\end{equation}
where $\mathrm{d}u/\mathrm{d}\fdm$ is the Jacobian that relates the observed fraction $u$ of lensed events to the compact dark matter fraction $\fdm$ in the Universe. 

We determine this Jacobian by simulating
astrophysical populations of \ac{BBH} mergers lensed by point mass lenses ~\citep{Basak:2021ten}.\footnote{
	The simulations are done assuming the O3b representative \ac{PSD} and Gaussian noise.
	The Jacobian is not expected to change significantly if real noise is used instead.
}
The constraints we obtain depend upon the assumed distributions of the component masses, spins and the redshifts of the mergers, which have considerable uncertainties.
We assume that the masses are distributed according to the \textsc{Power-law + Peak} model of \citet{LIGOScientific:2021psn} while spins are assumed to be aligned/antialigned with the orbital angular momentum with magnitudes distributed uniformly in (0, 0.99). We use the approximant \textsc{IMRPhenomD} \citep{Khan_2016IMRPhenomD} to produce the waveforms.
We consider different redshift distributions of the mergers: uniform distribution in comoving volume, the power-law model of \citet{LIGOScientific:2021psn}, the Madau-Dickinson model \citep{Madau:2014bja}, as well as some representative population-synthesis models given by \citet{Dominik:2013} and \citet{1602.04531}.
In our simulations, compact objects are approximated by point mass lenses and distributed uniformly in comoving volume.
Binaries producing a network \ac{SNR} of 8 or above in the LIGO--Virgo detectors are deemed detectable.
In order to reduce the computational cost of performing the simulations, we estimate $\BMLU$ using an approximation to the Bayes factor that is expected to be accurate in the high-\ac{SNR} regime~\citep{Cornish:2011ys,Vallisneri:2012qq}.
We then compute the fraction of detected events that produce a $\BMLU$ larger than the highest $\BMLU$ obtained from real LIGO--Virgo events.
This lensing fraction is computed as a function of the $\fdm$, which is used to compute the Jacobian $\mathrm{d}u/\mathrm{d}\fdm$.

\begin{figure}[tbh]
	\centering
	\includegraphics[width=\columnwidth]{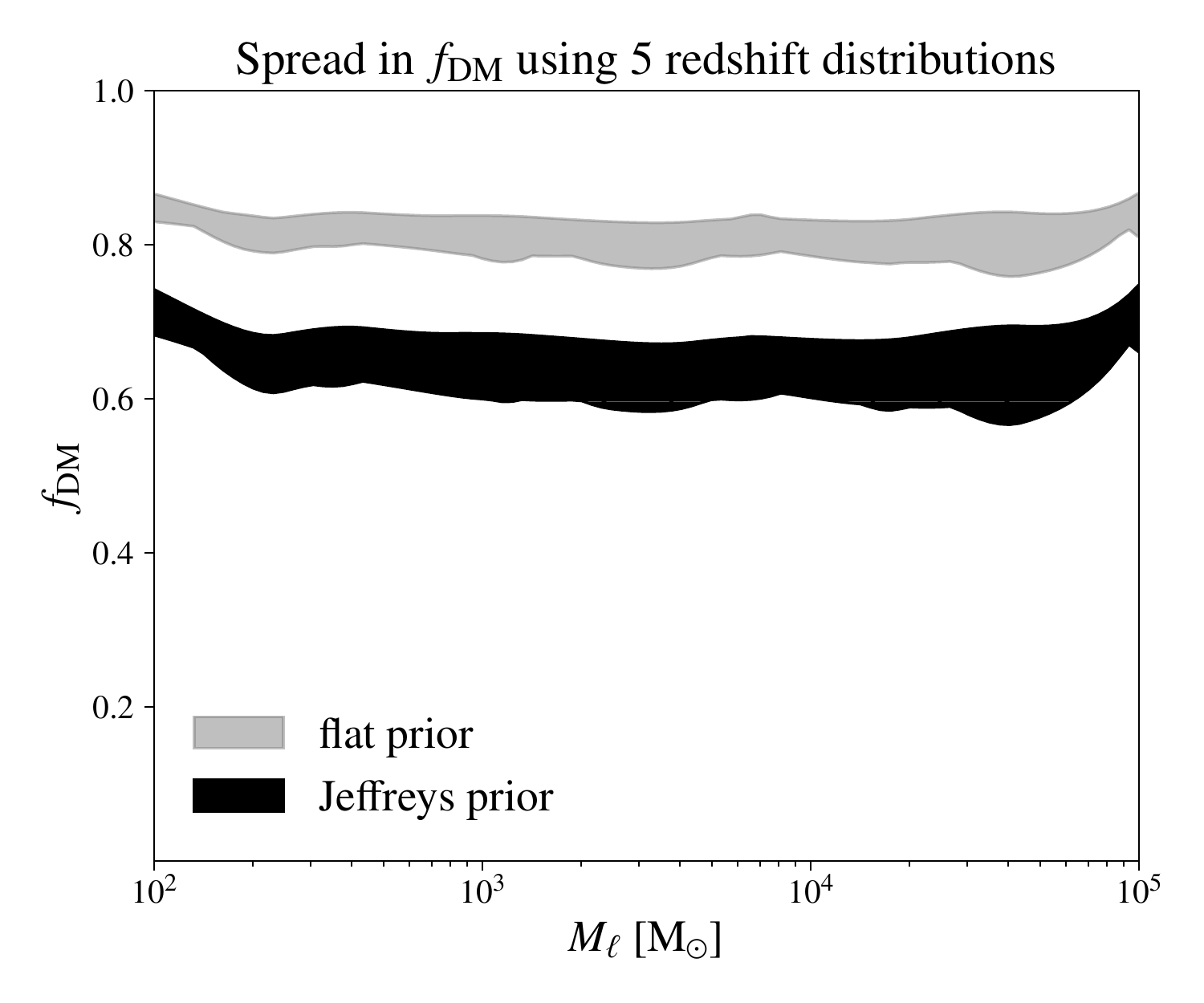}
	\caption{
		The spread in the 90\% upper limits on $\fdm$ obtained from the O3 events using 5 different redshift distribution models for \ac{BBH} mergers: \citet{1602.04531},\citet{Dominik:2013}, \citet{Madau:2014bja}, \citet{LIGOScientific:2021psn} and uniform in comoving 4-volume, assuming a monochromatic mass spectrum for the compact objects forming dark matter.
		The lens mass is shown on the horizontal axis.
		The grey (black) shaded regions correspond to the spread in $\fdm$ upper bounds computed assuming flat (Jeffreys) prior on $\Lambda$ and $\LambdaL$.
		The upper and lower curves bounding the spreads correspond to the most pessimistic (weakest) and optimistic (strongest) upper limits, as determined from the set of assumed redshift distributions, in each mass bin.
	}
	\label{fig:spread_in_fdm_upperlimits}
\end{figure}

The largest value of the microlensing likelihood ratio obtained from GWTC-3 events is $\log_{10}\BMLU = 0.799$. 
We compute the fraction of simulated events with $\log_{10}\BMLU \geq 0.799$, for different lens masses.
This allows us to compute the Jacobian $\mathrm{d}u/\mathrm{d}\fdm$ and thus the posterior on $\fdm$. 
The 90\% upper limits are shown as a function of the lens mass (assuming a monochromatic spectrum) in Fig.~\ref{fig:spread_in_fdm_upperlimits}. 
The bounds we obtain are  weaker than some of the existing constraints~\citep{Carr:2020xqk,Carr:2020gox}. 
The \ac{GW} lensing bounds will improve significantly in the next few years as the sensitivity of \ac{GW} detectors improve \citep{KAGRA:2013rdx}.
Assuming $\sim 300$ \ac{BBH} detections in O4 and $\mathcal{O}(1000)$ detections in O5, the constraints on $\fdm$ will improve to $\sim 10^{-1}$ and $\sim 10^{-2}$, respectively.

\section{Concluding Remarks}
\label{sec:conclusion}
We have extended the search for lensing signatures to all \ac{BBH} candidates with a probability of astrophysical origin higher than 0.5 from O3b~\citep{LIGOScientific:2021djp}.
While we have not observed any significant candidates for strongly lensed events, we updated the constraints on the rate of such events from several different analyses.
First, we searched for sub-threshold repeated signals associated with super-threshold events using reduced template banks produced from the posterior probability distributions of the super-threshold events.
Interesting sub-threshold/super-threshold pairs and pairs formed from two super-threshold events were further analyzed for their probability of being from a single, strongly lensed source.
For super-threshold/super-threshold pairs, we calculated the degree of overlap between the posteriors of the intrinsic parameters and sky location, which were obtained from Bayesian inference.
Moreover, we analyzed these pairs using a new analysis based on the comparison of spectrograms through machine learning.
Finally, pairs with false-positive probability from either analysis smaller than $10^{-2}$ were further studied by conducting full joint Bayesian inference analyses that take population priors and selection effects into account.
We found no pairs that show significant evidence for strong lensing.

The events from O3b were also analyzed for distortions caused by the lens on the gravitational waveform.
First, we searched for the distortions that lensing introduces on type II signals, which are in the form of a frequency-independent phase shift (Morse phase).
The Bayes factors for all events show no evidence for type II signal distortions.
Similarly, we searched for the frequency-dependent distortions caused by point masses.
None of the computed Bayes factors show any significant signs of microlensing.
For both analyses, some events show interesting features in the posteriors for the Morse phase or lens mass.
However, follow-up analyses using simulated signals show no further signs of the lensing nature of these features.
Altogether, we found no significant evidence for distortions of the gravitational waveforms that can be attributed to lensing.

The lack of evidence for lensing is then used to infer properties of the lensing rates and to set constraints on the dark matter fraction of (dark) compact objects.

Finally, we note that our conclusions are based on estimates and assumptions that are in line with other analyses from the LIGO--Virgo--KAGRA Collaboration \citep{GWTC3,LIGOScientific:2021psn}.
It is possible to arrive at different conclusions and interpretations if assumptions are chosen differently.
Examples include claims that almost all detections are strongly lensed if one assumes that heavy \acp{BH} do not exist \citep{Broadhurst:2018saj,Broadhurst:2020moy,Broadhurst:2020cvm}.
Data from the upcoming observing runs are expected to further expand the catalog of \ac{GW} detections that can further shed light on the lensing of \acp{GW} \citep{Aasi:2013wya}.
Moreover, multi-messenger astronomy may provide significant input in confirming and interpreting possible lensed \ac{GW} signals \citep{Wempe:2022zlk}.

\acknowledgments
{
	This material is based upon work supported by NSF’s LIGO Laboratory which is a major facility
fully funded by the National Science Foundation.
The authors also gratefully acknowledge the support of
the Science and Technology Facilities Council (STFC) of the
United Kingdom, the Max-Planck-Society (MPS), and the State of
Niedersachsen/Germany for support of the construction of Advanced LIGO 
and construction and operation of the GEO\,600 detector. 
Additional support for Advanced LIGO was provided by the Australian Research Council.
The authors gratefully acknowledge the Italian Istituto Nazionale di Fisica Nucleare (INFN),  
the French Centre National de la Recherche Scientifique (CNRS) and
the Netherlands Organization for Scientific Research (NWO), 
for the construction and operation of the Virgo detector
and the creation and support  of the EGO consortium. 
The authors also gratefully acknowledge research support from these agencies as well as by 
the Council of Scientific and Industrial Research of India, 
the Department of Science and Technology, India,
the Science \& Engineering Research Board (SERB), India,
the Ministry of Human Resource Development, India,
the Spanish Agencia Estatal de Investigaci\'on (AEI),
the Spanish Ministerio de Ciencia e Innovaci\'on and Ministerio de Universidades,
the Conselleria de Fons Europeus, Universitat i Cultura and the Direcci\'o General de Pol\'{\i}tica Universitaria i Recerca del Govern de les Illes Balears,
the Conselleria d'Innovaci\'o, Universitats, Ci\`encia i Societat Digital de la Generalitat Valenciana and
the CERCA Programme Generalitat de Catalunya, Spain,
the National Science Centre of Poland and the European Union – European Regional Development Fund; Foundation for Polish Science (FNP),
the Swiss National Science Foundation (SNSF),
the Russian Foundation for Basic Research, 
the Russian Science Foundation,
the European Commission,
the European Social Funds (ESF),
the European Regional Development Funds (ERDF),
the Royal Society, 
the Scottish Funding Council, 
the Scottish Universities Physics Alliance, 
the Hungarian Scientific Research Fund (OTKA),
the French Lyon Institute of Origins (LIO),
the Belgian Fonds de la Recherche Scientifique (FRS-FNRS), 
Actions de Recherche Concertées (ARC) and
Fonds Wetenschappelijk Onderzoek – Vlaanderen (FWO), Belgium,
the Paris \^{I}le-de-France Region, 
the National Research, Development and Innovation Office Hungary (NKFIH), 
the National Research Foundation of Korea,
the Natural Science and Engineering Research Council Canada,
Canadian Foundation for Innovation (CFI),
the Brazilian Ministry of Science, Technology, and Innovations,
the International Center for Theoretical Physics South American Institute for Fundamental Research (ICTP-SAIFR), 
the Research Grants Council of Hong Kong,
the National Natural Science Foundation of China (NSFC),
the Leverhulme Trust, 
the Research Corporation,
the National Science and Technology Council (NSTC), Taiwan,
the United States Department of Energy,
and
the Kavli Foundation.
The authors gratefully acknowledge the support of the NSF, STFC, INFN and CNRS for provision of computational resources.

This work was supported by MEXT, JSPS Leading-edge Research Infrastructure Program, JSPS Grant-in-Aid for Specially Promoted Research 26000005, JSPS Grant-inAid for Scientific Research on Innovative Areas 2905: JP17H06358, JP17H06361 and JP17H06364, JSPS Core-to-Core Program A. Advanced Research Networks, JSPS Grantin-Aid for Scientific Research (S) 17H06133 and 20H05639 , JSPS Grant-in-Aid for Transformative Research Areas (A) 20A203: JP20H05854, the joint research program of the Institute for Cosmic Ray Research, University of Tokyo, National Research Foundation (NRF), Computing Infrastructure Project of Global Science experimental Data hub Center (GSDC) at KISTI, Korea Astronomy and Space Science Institute (KASI), and Ministry of Science and ICT (MSIT) in Korea, Academia Sinica (AS), AS Grid Center (ASGC) and the National Science and Technology Council (NSTC) in Taiwan under grants including the Rising Star Program and Science Vanguard Research Program, Advanced Technology Center (ATC) of NAOJ, and Mechanical Engineering Center of KEK.

\software{
Analyses in this paper made use of 
\texttt{LALSuite}~\citep{lalsuite},
the
\texttt{GstLAL}~\citep{Cannon:2011vi, Messick:2016aqy, Hanna:2019ezx, Sachdev:2019vvd} pipeline;
Bayesian inference with
\texttt{bilby}~\citep{Ashton:2018jfp, Smith:2019ucc,Romero-Shaw:2020owr};
as well as the packages
\texttt{NumPy} \citep{Harris:2020xlr},
\texttt{SciPy} \citep{2020SciPy-NMeth},
\texttt{Astropy} \citep{astropy:2013,astropy:2018},
\texttt{IPython} \citep{ipython}, 
and \texttt{ligo.skymap} \citep{ligoskymap}.
Plots were produced with
\texttt{Matplotlib} \citep{matplotlib},
and \texttt{Seaborn} \citep{seaborn}.
}

}

\vspace{5mm}

\appendix

\section{Type II simulation campaigns} 
\label{sec:app_typeII}

Given the mild evidence of GW190412 and GW200129\_065458 towards being a type II strongly lensed image presented in Sec. \ref{sec:typeII}, we follow up on these events by doing an injection campaign where we simulate type I and type II images similar to the events, and verify whether the posteriors recovered are compatible with the distribution observed for the real events. 
These injections are performed in different noise realizations and with different waveform models. 
The observed feature could be caused by two main other effects than a type II image: noise artifacts or systematic effects in the waveform modeling. 
In the former, non-Gaussianities in the noise could be such that they lead to the observation of spurious features, while in the latter case, the specific combination of observed parameters could lead to some systematic issues in fitting with the waveform model.
Waveform systematics might be especially important for these events since they lie in challenging parts of the parameter space~\citep{LIGOScientific:2020stg,Colleoni:2020tgc,Hannam:2021pit}. 
Moreover, for GW200129\_065458 \citet{Payne:2022spz} reports that there could be a significant glitch under subtraction.

\begin{figure}[h]
	\centering
	\includegraphics[width=0.6\columnwidth]{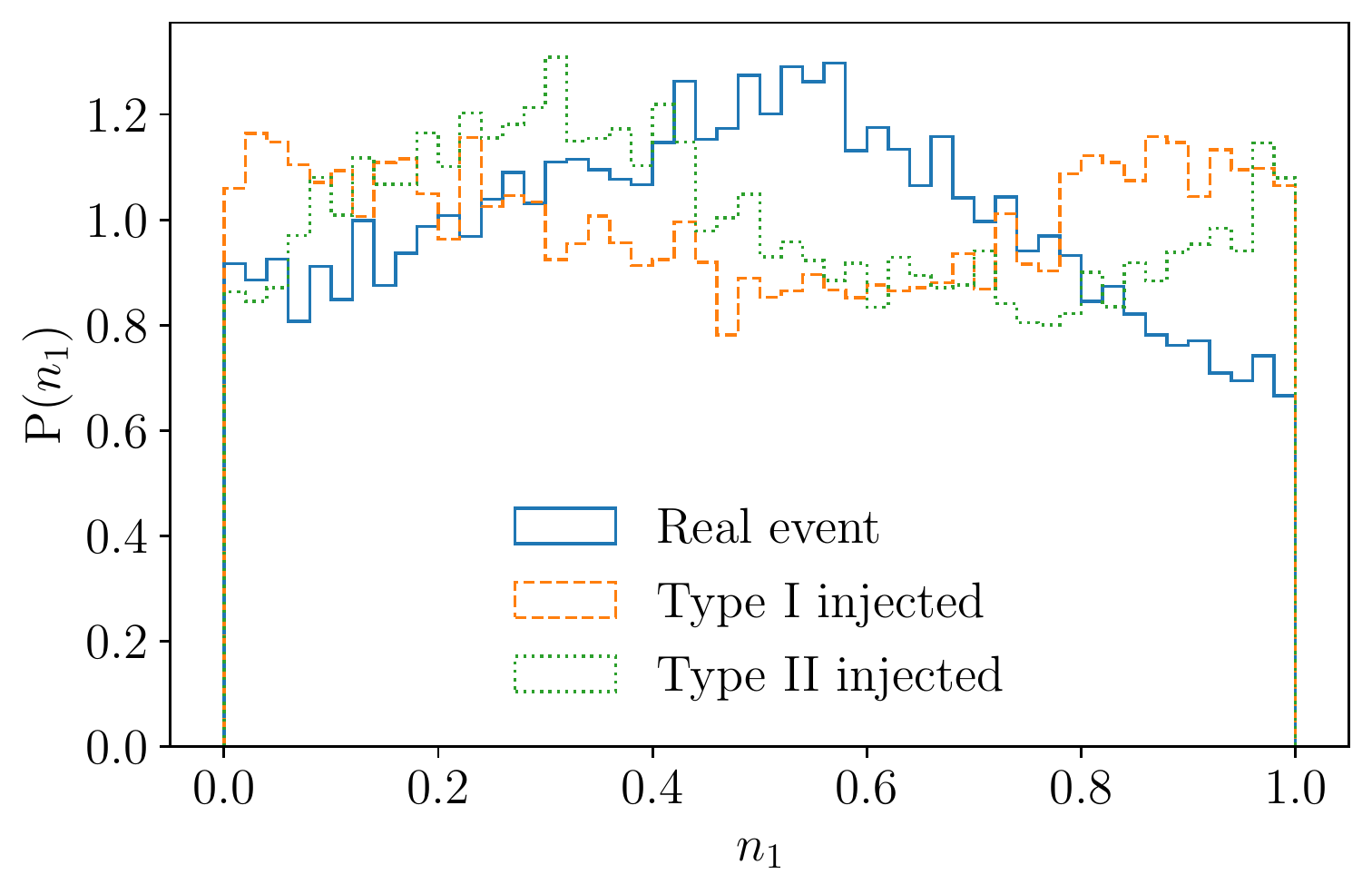}
	\caption{
		Posterior distribution of the Morse phase for GW200129\_065458. 
		We compare the real event posterior (solid-blue) with an injection campaign of type I (dashed-orange) and type II (dotted-green) images. 
		Type II images correspond to $n_1=1/2$.
		For this event, the differences between the distribution are small and make it difficult to learn anything additional about the event. 
		The Kolmogorov--Smirnov statistic is 0.07 for type I vs real, and 0.08 for type II vs real.
	}
	\label{fig:S200129m_Injection}
\end{figure}

\begin{figure}[h]
	\centering
	\includegraphics[width=0.6\columnwidth]{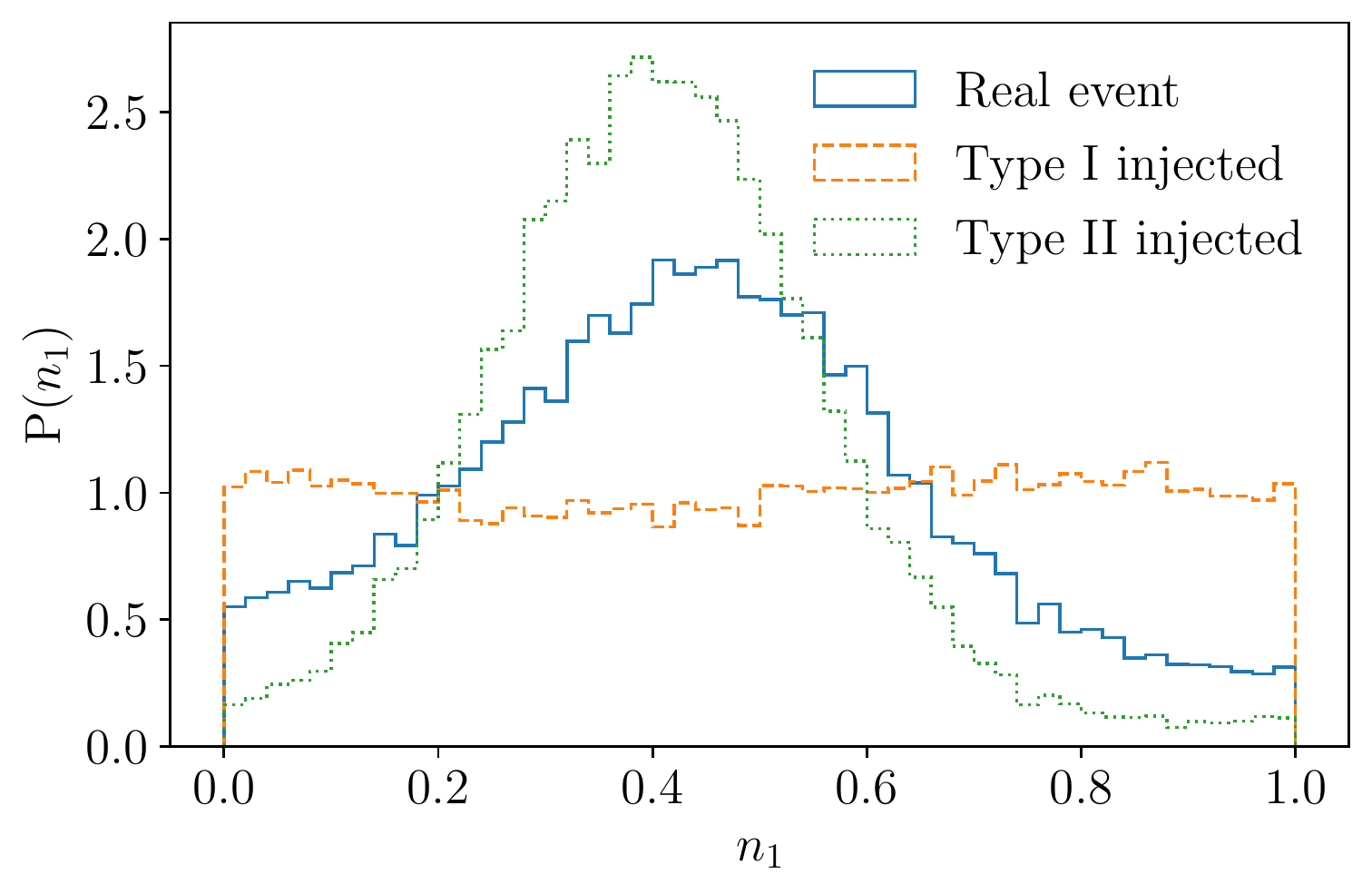}
	\caption{
		Posterior distribution of the Morse phase for GW190412. 
		We compare the real event posterior (solid-blue) with an injection campaign of type I (dashed-orange) and type II (dotted-green) images. 
		Type II images correspond to $n_1=1/2$. 
		For this event, the peak seen in the real data and the one seen for the type II image are compatible, hinting at a possible type II image. 
		In this case, the Kolmogorov--Smirnov statistic is 0.20 for type I vs real, and 0.13 for type II vs real.
	}
	\label{fig:S190412m_InjectionFirst}
\end{figure}

To test for the noise-related features, we generate colored Gaussian noise from the PSD around the time of the candidate and then inject the maximum likelihood parameters coming from the parameter estimation and take the Morse factor to be either the value for a type I or a type II image.
In the first step, we do this for only one noise realization for each event to see whether we can reproduce similar features or not.
For GW200129\_065458, the injection shows that the effect is too weak to be distinguishable from one image type to the other, as can be seen in the uninformative posteriors of Fig.~\ref{fig:S200129m_Injection}.
As a consequence, no further investigation is done into this event.
On the other hand, for GW190412, the feature seen in the real data is compatible with the one seen in the injection (see Fig.~\ref{fig:S190412m_InjectionFirst}).

Given that the real-data results are compatible with the type II injection for GW190412, we investigate further the noise hypothesis. 
For this purpose, we take the maximum likelihood parameters and a Morse factor of 0 or $1/2$ and inject the signal generated with the \IMRXPHM~\citep{Pratten:2020ceb} model in ten different noise realizations. 
We then repeat the analysis in the same way as for the real signal and verify if we retrieve the same preference for a type II image. 
For all the noise realizations used here, we see the same behavior as in Fig.~\ref{fig:S190412m_InjectionFirst}.

\begin{figure}[h]
	\centering
	\includegraphics[width=0.6\columnwidth]{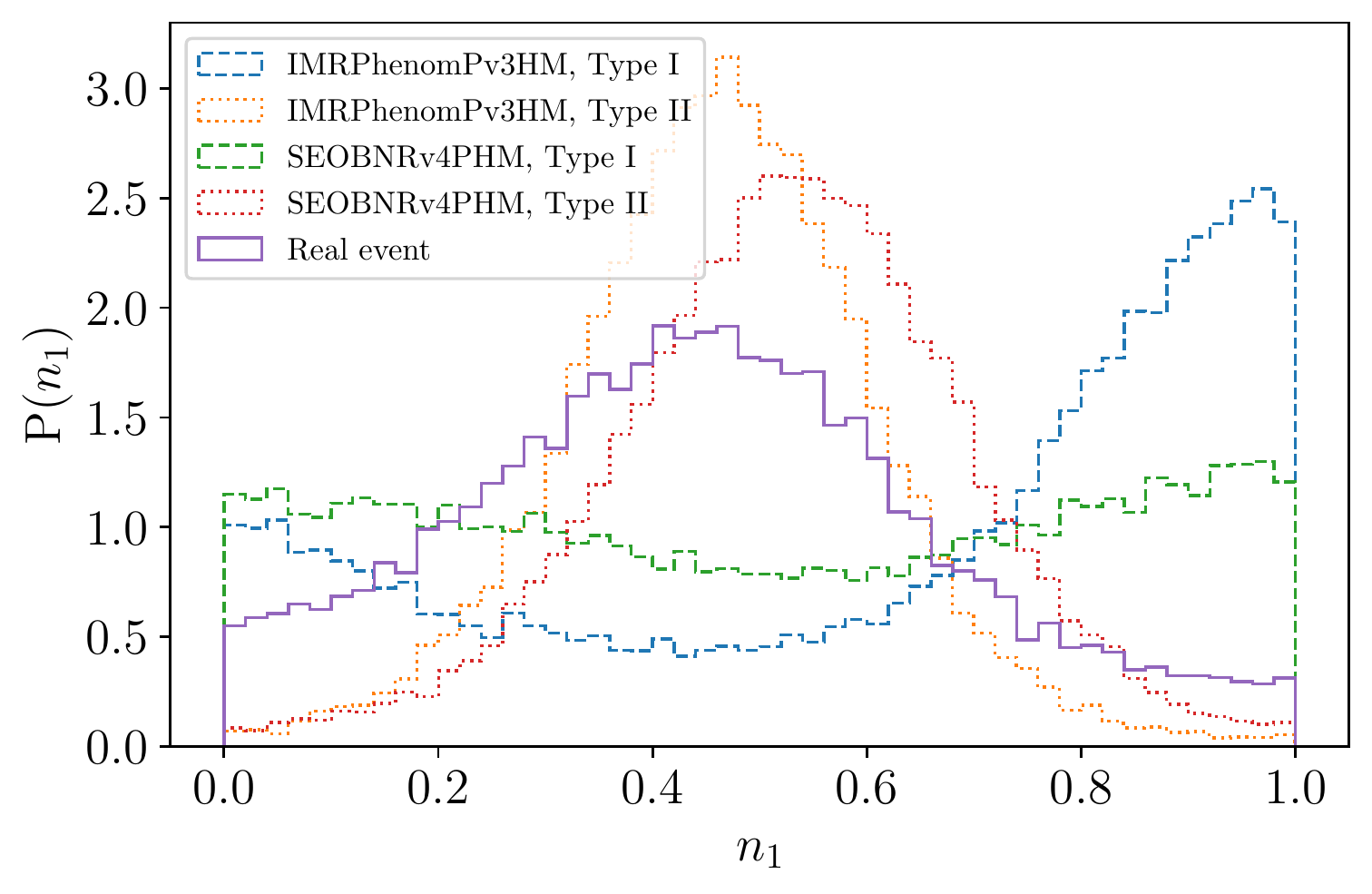}
	\caption{
		Comparison of the Morse factor distribution for the real event (solid-purple) with the recovered posterior distribution for an injection made with \textsc{IMRPhenomPv3HM} for a type I image (dashed-blue) and a type II image (dotted-orange), and with an injection made with \textsc{SEOBNRv4PHM} for a type I image (dashed-green) and a type II image (dotted-red).
		In all the cases, the posterior distributions agree with the injected data, with the real event resembling a type II image.
	}
	\label{fig:OtherWaveformsTypeII}
\end{figure}

We perform an extra test by injecting the maximum likelihood parameters with a given image type in the generated noise for different waveform models.
We use the \textsc{IMRPhenomPv3HM}~\citep{Khan:2019kot} and the \textsc{SEOBNRv4PHM}~\citep{Ossokine:2020kjp} model to generate the signal and use the \IMRXPHM~\citep{Pratten:2020ceb} model to recover it. 
This enables us to combine the two possible sources of systematics. 
This way, we can verify whether a different noise combined with a different model also leads to a preference for type II images. 
For all the noise realizations and the two models used for the injections, we find that the injections always recover the correct hypothesis, and the fact that the real event supports type II is unlikely to be a result of noise or waveform artifacts, as shown in Fig.~\ref{fig:OtherWaveformsTypeII}. 

Although these tests do not discard the type II image hypothesis, they cannot conclusively confirm it. 
To confirm the presence of lensing for this event with a mild preference for a type II image, we would need additional evidence. 
Therefore, we search for possible sub-threshold counterparts with the methodology explained in Sec.~\ref{sec:subthreshold}. 
However, we find only marginal triggers.

In the end, these additional searches did not enable us to find any extra evidence for lensing, while still not ruling out the possibility for GW190412 to be a type II image.

\section{Marginalized Posteriors of Microlensing Parameters}
\label{sec:app_micro}

\begin{figure}[h!]
	\centering
	\includegraphics[width=0.75\columnwidth]{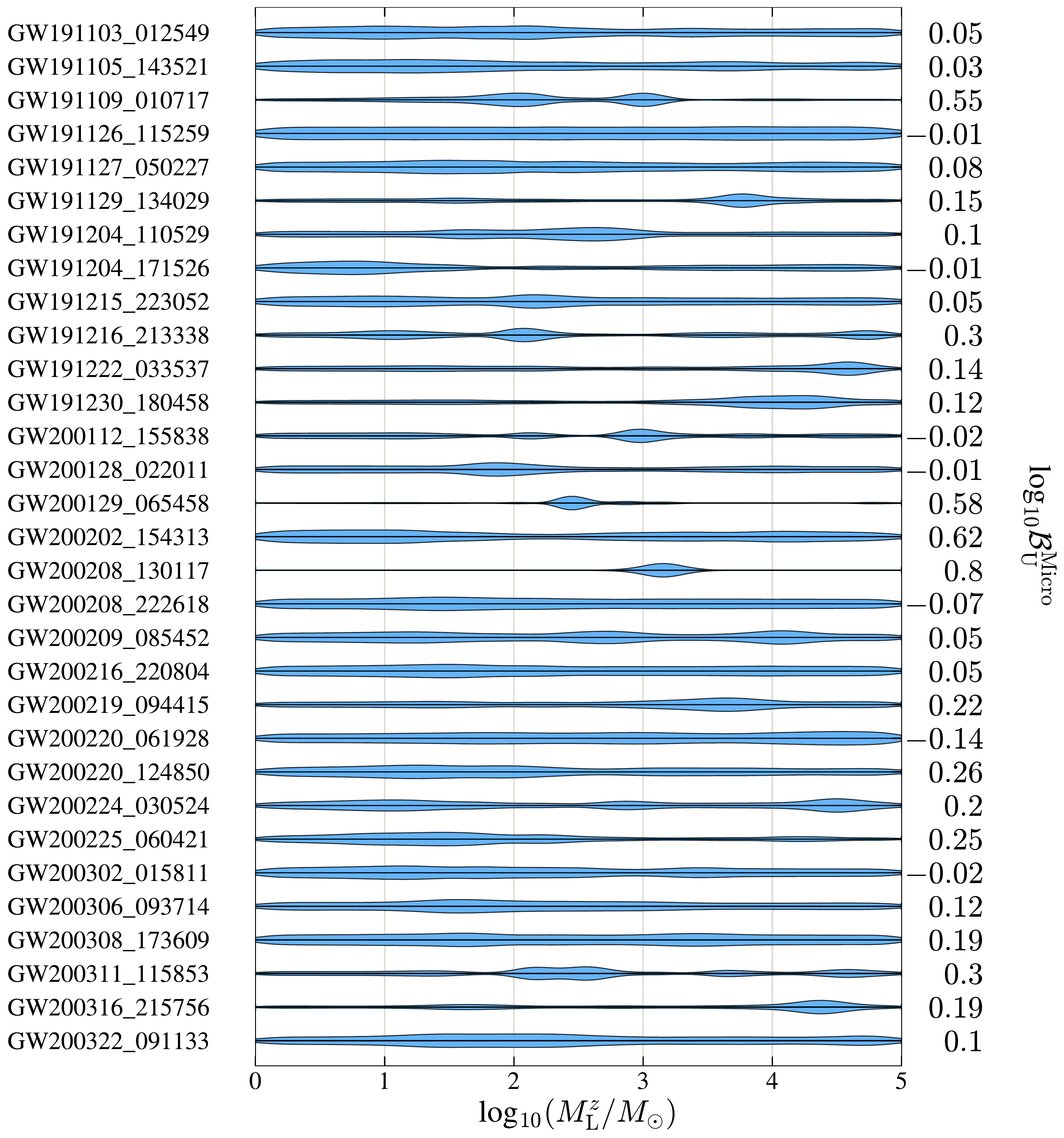}
	\caption{Marginalized posterior distributions of redshifted lens mass $\MzL$ and $\log_{10}\BMLU$ between microlensed and unlensed hypotheses.}
	\label{fig:micro_violin}
\end{figure}

\begin{figure}[h!]
	\centering
	\includegraphics[width=0.6\columnwidth]{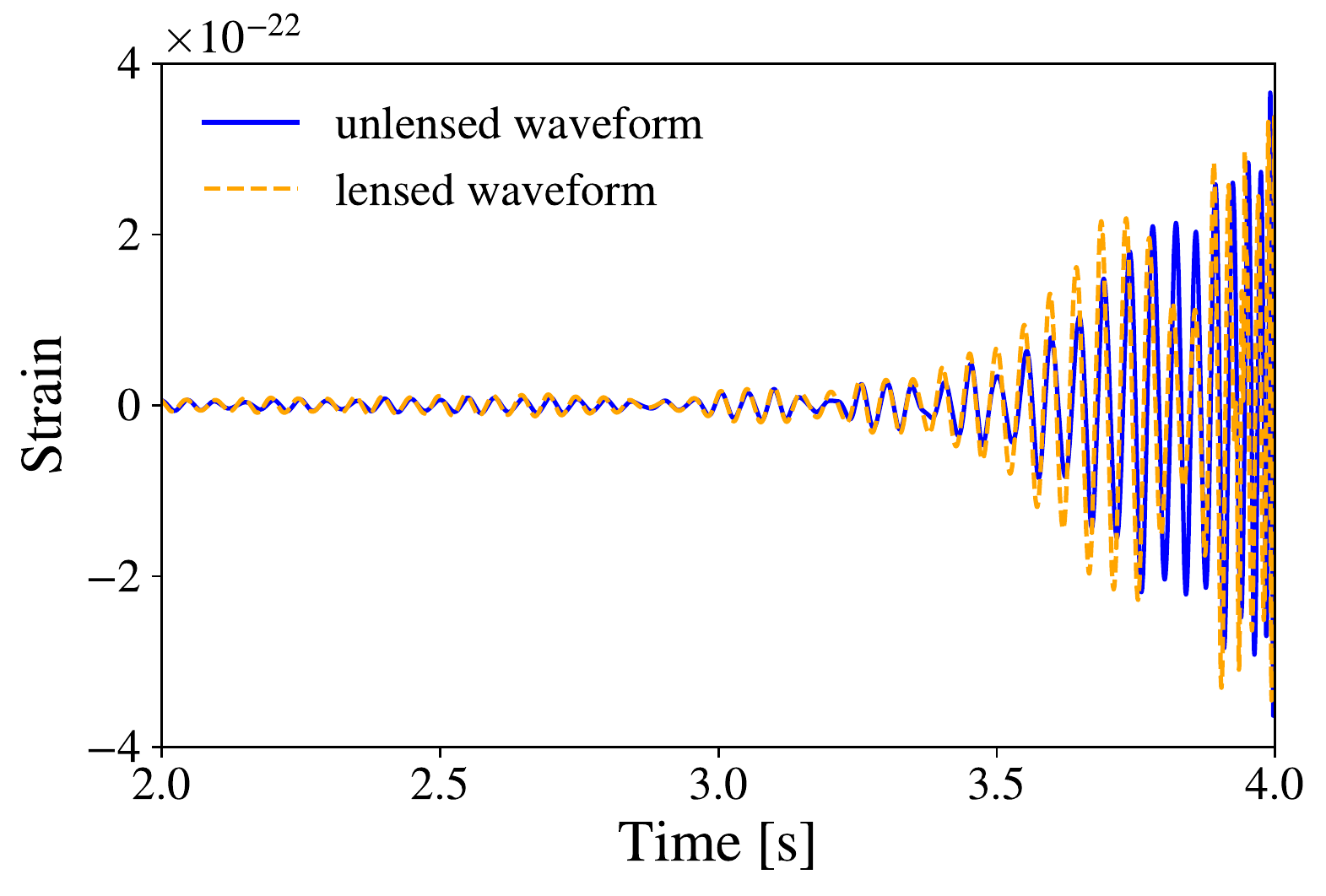}
	\caption{The time-domain waveform corresponding to the maximum posterior of GW200208\_130117, with and without the microlensing hypothesis.}
	\label{fig:micro_waveform}
\end{figure}

As a supplement to the distribution of $\log_{10}$~Bayes factors $\BMLU$ shown in Fig.~\ref{fig:micro_BLUdistr}, we show the individual marginalized posterior distributions of redshifted lens mass $\MzL$ and $\log_{10}\BMLU$ (right vertical axis) in Fig.~\ref{fig:micro_violin}.
The Bayes factors individually do not show clear evidence for microlensing by point-mass lenses.
However, several events show a narrow posterior distribution of the redshifted lens mass.
An example is GW200208\_130117 (with $\log_{10}\BMLU = 0.8$), for which the waveform corresponding to the maximum posterior for this event, with and without lensing, is shown in Fig.~\ref{fig:micro_waveform}.
The beating pattern introduced by the point-mass lens is most visible as a reduction of the amplitude for two cycles in the middle of the signal and an increase in the amplitude before and after this reduction.
We hypothesize that short-duration noise fluctuations may have caused an apparent dip in the signal, which in turn may have led to a distortion similar to a point-mass lens beating pattern.
This is corroborated by a low Bayes factor $\BMLU$, which concludes the data is inconclusive about the microlensing hypothesis.

\bibliographystyle{aasjournal}
\bibliography{cbc-group,software,lensing}

\begin{thebibliography}{}
\expandafter\ifx\csname natexlab\endcsname\relax\def\natexlab#1{#1}\fi
\providecommand{\url}[1]{\href{#1}{#1}}
\providecommand{\dodoi}[1]{doi:~\href{http://doi.org/#1}{\nolinkurl{#1}}}
\providecommand{\doeprint}[1]{\href{http://ascl.net/#1}{\nolinkurl{http://ascl.net/#1}}}
\providecommand{\doarXiv}[1]{\href{https://arxiv.org/abs/#1}{\nolinkurl{https://arxiv.org/abs/#1}}}

\bibitem[{Abbott {et~al.}(2016{\natexlab{a}})}]{TheLIGOScientific:2016zmo}
Abbott, B.~P., {et~al.} 2016{\natexlab{a}}, Class. Quant. Grav., 33, 134001,
  \dodoi{10.1088/0264-9381/33/13/134001}

\bibitem[{Abbott {et~al.}(2016{\natexlab{b}})}]{Abbott:2016nhf}
---. 2016{\natexlab{b}}, Astrophys. J., 833, L1,
  \dodoi{10.3847/2041-8205/833/1/L1}

\bibitem[{Abbott {et~al.}(2018)}]{KAGRA:2013rdx}
---. 2018, Living Rev. Rel., 21, 3, \dodoi{10.1007/s41114-020-00026-9}

\bibitem[{Abbott {et~al.}(2020{\natexlab{a}})}]{LIGOScientific:2019hgc}
---. 2020{\natexlab{a}}, Class. Quant. Grav., 37, 055002,
  \dodoi{10.1088/1361-6382/ab685e}

\bibitem[{Abbott {et~al.}(2020{\natexlab{b}})}]{Aasi:2013wya}
---. 2020{\natexlab{b}}, Living Rev. Relativity, 23, 3,
  \dodoi{10.1007/s41114-020-00026-9}

\bibitem[{Abbott {et~al.}(2020{\natexlab{c}})}]{LIGOScientific:2020stg}
Abbott, R., {et~al.} 2020{\natexlab{c}}, Phys. Rev. D, 102, 043015,
  \dodoi{10.1103/PhysRevD.102.043015}

\bibitem[{Abbott {et~al.}(2020{\natexlab{d}})}]{GW190814}
---. 2020{\natexlab{d}}, \dodoi{10.3847/2041-8213/ab960f}

\bibitem[{Abbott {et~al.}(2021{\natexlab{a}})}]{LIGOScientific:2021izm}
---. 2021{\natexlab{a}}, Astrophys. J., 923, 14,
  \dodoi{10.3847/1538-4357/ac23db}

\bibitem[{Abbott {et~al.}(2021{\natexlab{b}})}]{LIGOScientific:2021djp}
---. 2021{\natexlab{b}}.
\newblock \doarXiv{2111.03606}

\bibitem[{Abbott {et~al.}(2021{\natexlab{c}})}]{datarelease}
---. 2021{\natexlab{c}}, Data release for "Search for gravitational-lensing
  signatures in the full third observing run of the LIGO-Virgo network".
\newblock \url{https://dcc.ligo.org/XXXXXXXX/public}

\bibitem[{Abbott
  {et~al.}(2021{\natexlab{d}})}]{ligo_scientific_collaboration_and_virgo_2021_5546663}
---. 2021{\natexlab{d}}, {GWTC-3: Compact Binary Coalescences Observed by LIGO
  and Virgo During the Second Part of the Third Observing Run — Parameter
  estimation data release},  Zenodo, \dodoi{10.5281/zenodo.5546663}

\bibitem[{Abbott {et~al.}(2021{\natexlab{e}})}]{Abbott:2019ebz}
---. 2021{\natexlab{e}}, SoftwareX, 100658, \dodoi{10.1016/j.softx.2021.100658}

\bibitem[{Abbott {et~al.}(2021{\natexlab{f}})}]{GWTC3}
---. 2021{\natexlab{f}}.
\newblock \doarXiv{2111.03606}

\bibitem[{Abbott {et~al.}(2021{\natexlab{g}})}]{GWTC2}
---. 2021{\natexlab{g}}, Phys. Rev. X, 11, 021053,
  \dodoi{10.1103/PhysRevX.11.021053}

\bibitem[{Abbott {et~al.}(2021{\natexlab{h}})}]{GWTC-2.1}
---. 2021{\natexlab{h}}.
\newblock \doarXiv{2108.01045}

\bibitem[{Abbott {et~al.}(2021{\natexlab{i}})}]{LIGOScientific:2021qlt}
---. 2021{\natexlab{i}}, Astrophys. J. Lett., 915, L5,
  \dodoi{10.3847/2041-8213/ac082e}

\bibitem[{Abbott {et~al.}(2021{\natexlab{j}})}]{LIGOScientific:2021psn}
---. 2021{\natexlab{j}}.
\newblock \doarXiv{2111.03634}

\bibitem[{Abbott {et~al.}(2021{\natexlab{k}})}]{Abbott:2021xxi}
---. 2021{\natexlab{k}}.
\newblock \doarXiv{2101.12130}

\bibitem[{Abbott {et~al.}(2021{\natexlab{l}})}]{KAGRA:2021kbb}
---. 2021{\natexlab{l}}, Phys. Rev. D, 104, 022004,
  \dodoi{10.1103/PhysRevD.104.022004}

\bibitem[{Abbott {et~al.}(2021{\natexlab{m}})}]{Abbott:2020gyp}
---. 2021{\natexlab{m}}, Astrophys. J. Lett., 913, L7,
  \dodoi{10.3847/2041-8213/abe949}

\bibitem[{Acernese {et~al.}(2021)}]{Acernese2021}
Acernese, F., {et~al.} 2021.
\newblock \doarXiv{2107.03294}

\bibitem[{Acernese {et~al.}(2022)}]{Acernese:2022jes}
---. 2022.
\newblock \doarXiv{2205.01555}

\bibitem[{Adams {et~al.}(2016)Adams, Buskulic, Germain, Guidi, Marion, Montani,
  Mours, Piergiovanni, \& Wang}]{Adams_2016}
Adams, T., Buskulic, D., Germain, V., {et~al.} 2016, Classical and Quantum
  Gravity, 33, 175012, \dodoi{10.1088/0264-9381/33/17/175012}

\bibitem[{{Ade} {et~al.}(2016)}]{Planck_2015}
{Ade}, P.~A.~R., {et~al.} 2016, \aap, 594, A13,
  \dodoi{10.1051/0004-6361/201525830}

\bibitem[{Allen(2005)}]{Allen:2004gu}
Allen, B. 2005, Phys. Rev. D, 71, 062001, \dodoi{10.1103/PhysRevD.71.062001}

\bibitem[{Allen {et~al.}(2012{\natexlab{a}})Allen, Anderson, Brady, Brown, \&
  Creighton}]{PhysRevD.85.122006}
Allen, B., Anderson, W.~G., Brady, P.~R., Brown, D.~A., \& Creighton, J. D.~E.
  2012{\natexlab{a}}, Phys. Rev. D, 85, 122006,
  \dodoi{10.1103/PhysRevD.85.122006}

\bibitem[{Allen {et~al.}(2012{\natexlab{b}})Allen, Anderson, Brady, Brown, \&
  Creighton}]{Allen:2005fk}
---. 2012{\natexlab{b}}, Phys. Rev. D, 85, 122006,
  \dodoi{10.1103/PhysRevD.85.122006}

\bibitem[{Ashton {et~al.}(2019)}]{Ashton:2018jfp}
Ashton, G., {et~al.} 2019, Astrophys. J. Suppl. Ser., 241, 27,
  \dodoi{10.3847/1538-4365/ab06fc}

\bibitem[{Aubin {et~al.}(2021)Aubin, Brighenti, Chierici, Estevez, Greco,
  Guidi, Juste, Marion, Mours, Nitoglia, Sauter, \& Sordini}]{Aubin_2021}
Aubin, F., Brighenti, F., Chierici, R., {et~al.} 2021, Classical and Quantum
  Gravity, 38, 095004, \dodoi{10.1088/1361-6382/abe913}

\bibitem[{Baker \& Trodden(2017)}]{Baker:2016reh}
Baker, T., \& Trodden, M. 2017, Phys. Rev. D, 95, 063512,
  \dodoi{10.1103/PhysRevD.95.063512}

\bibitem[{Basak {et~al.}(2021)Basak, Ganguly, Haris, Kapadia, Mehta, \&
  Ajith}]{Basak:2021ten}
Basak, S., Ganguly, A., Haris, K., {et~al.} 2021.
\newblock \doarXiv{2109.06456}

\bibitem[{Belczynski {et~al.}(2010)Belczynski, Dominik, Bulik, O'Shaughnessy,
  Fryer, \& Holz}]{Belczynski:2010tb}
Belczynski, K., Dominik, M., Bulik, T., {et~al.} 2010, Astrophys. J. Lett.,
  715, L138, \dodoi{10.1088/2041-8205/715/2/L138}

\bibitem[{Belczynski {et~al.}(2016)Belczynski, Holz, Bulik, \&
  O’Shaughnessy}]{1602.04531}
Belczynski, K., Holz, D.~E., Bulik, T., \& O’Shaughnessy, R. 2016, Nature,
  534, 512–515, \dodoi{10.1038/nature18322}

\bibitem[{Belczynski {et~al.}(2008)Belczynski, Kalogera, Rasio, Taam, Zezas,
  Bulik, Maccarone, \& Ivanova}]{Belczynski:2005mr}
Belczynski, K., Kalogera, V., Rasio, F.~A., {et~al.} 2008, Astrophys. J.
  Suppl., 174, 223, \dodoi{10.1086/521026}

\bibitem[{Blanchet(2014)}]{Blanchet:2013haa}
Blanchet, L. 2014, Living Rev. Relativity, 17, 2, \dodoi{10.12942/lrr-2014-2}

\bibitem[{Bouffanais {et~al.}(2021)Bouffanais, Mapelli, Santoliquido, Giacobbo,
  Di~Carlo, Rastello, Artale, \& Iorio}]{Bouffanais:2021wcr}
Bouffanais, Y., Mapelli, M., Santoliquido, F., {et~al.} 2021.
\newblock \doarXiv{2102.12495}

\bibitem[{Broadhurst {et~al.}(2018)Broadhurst, Diego, \&
  Smoot}]{Broadhurst:2018saj}
Broadhurst, T., Diego, J.~M., \& Smoot, G. 2018.
\newblock \doarXiv{1802.05273}

\bibitem[{Broadhurst {et~al.}(2020{\natexlab{a}})Broadhurst, Diego, \&
  Smoot}]{Broadhurst:2020moy}
Broadhurst, T., Diego, J.~M., \& Smoot, G.~F. 2020{\natexlab{a}}.
\newblock \doarXiv{2002.08821}

\bibitem[{Broadhurst {et~al.}(2020{\natexlab{b}})Broadhurst, Diego, \&
  Smoot}]{Broadhurst:2020cvm}
---. 2020{\natexlab{b}}.
\newblock \doarXiv{2006.13219}

\bibitem[{Buscicchio {et~al.}(2020)Buscicchio, Moore, Pratten, Schmidt,
  Bianconi, \& Vecchio}]{Buscicchio:2020cij}
Buscicchio, R., Moore, C.~J., Pratten, G., {et~al.} 2020, Phys. Rev. Lett.,
  125, 141102, \dodoi{10.1103/PhysRevLett.125.141102}

\bibitem[{Cahillane {et~al.}(2017)}]{Cahillane:2017vkb}
Cahillane, C., {et~al.} 2017, Phys. Rev. D, 96, 102001,
  \dodoi{10.1103/PhysRevD.96.102001}

\bibitem[{Cannon {et~al.}(2012)Cannon, Cariou, Chapman, Crispin-Ortuzar,
  Fotopoulos, {et~al.}}]{Cannon:2011vi}
Cannon, K., Cariou, R., Chapman, A., {et~al.} 2012, Astrophys.J., 748, 136,
  \dodoi{10.1088/0004-637X/748/2/136}

\bibitem[{Cao {et~al.}(2019)Cao, Qi, Cao, Biesiada, Li, Pan, \&
  Zhu}]{Cao:2019kgn}
Cao, S., Qi, J., Cao, Z., {et~al.} 2019, Sci. Rep., 9, 11608,
  \dodoi{10.1038/s41598-019-47616-4}

\bibitem[{Cao {et~al.}(2014)Cao, Li, \& Wang}]{Cao:2014oaa}
Cao, Z., Li, L.-F., \& Wang, Y. 2014, Phys. Rev. D, 90, 062003,
  \dodoi{10.1103/PhysRevD.90.062003}

\bibitem[{Carr {et~al.}(2020)Carr, Kohri, Sendouda, \& Yokoyama}]{Carr:2020gox}
Carr, B., Kohri, K., Sendouda, Y., \& Yokoyama, J. 2020.
\newblock \doarXiv{2002.12778}

\bibitem[{Carr \& Kuhnel(2020)}]{Carr:2020xqk}
Carr, B., \& Kuhnel, F. 2020, Ann. Rev. Nucl. Part. Sci., 70, 355,
  \dodoi{10.1146/annurev-nucl-050520-125911}

\bibitem[{\c{C}al\i{}\c{s}kan {et~al.}(2022{\natexlab{a}})\c{C}al\i{}\c{s}kan,
  Ezquiaga, Hannuksela, \& Holz}]{Caliskan:2022wbh}
\c{C}al\i{}\c{s}kan, M., Ezquiaga, J.~M., Hannuksela, O.~A., \& Holz, D.~E.
  2022{\natexlab{a}}.
\newblock \doarXiv{2201.04619}

\bibitem[{\c{C}al\i{}\c{s}kan {et~al.}(2022{\natexlab{b}})\c{C}al\i{}\c{s}kan,
  Ji, Cotesta, Berti, Kamionkowski, \& Marsat}]{Caliskan:2022hbu}
\c{C}al\i{}\c{s}kan, M., Ji, L., Cotesta, R., {et~al.} 2022{\natexlab{b}}.
\newblock \doarXiv{2206.02803}

\bibitem[{Chatterji {et~al.}(2004)Chatterji, Blackburn, Martin, \&
  Katsavounidis}]{Chatterji:2004qg}
Chatterji, S., Blackburn, L., Martin, G., \& Katsavounidis, E. 2004, Class.
  Quantum Grav., 21, S1809, \dodoi{10.1088/0264-9381/21/20/024}

\bibitem[{Chen \& Guestrin(2016)}]{Chen:2016btl}
Chen, T., \& Guestrin, C. 2016, in Proceedings of the 22nd ACM SIGKDD
  International Conference on Knowledge Discovery and Data Mining, KDD '16 (New
  York, NY, USA: Association for Computing Machinery), 785–794,
  \dodoi{10.1145/2939672.2939785}

\bibitem[{Cheung {et~al.}(2021)Cheung, Gais, Hannuksela, \&
  Li}]{Cheung:2020okf}
Cheung, M. H.~Y., Gais, J., Hannuksela, O.~A., \& Li, T. G.~F. 2021, Mon. Not.
  Roy. Astron. Soc., 503, 3326, \dodoi{10.1093/mnras/stab579}

\bibitem[{{Choi} {et~al.}(2007){Choi}, {Park}, \& {Vogeley}}]{Choi:2007a}
{Choi}, Y.-Y., {Park}, C., \& {Vogeley}, M.~S. 2007, Astrophys. J., 658, 884,
  \dodoi{10.1086/511060}

\bibitem[{Christian {et~al.}(2018)Christian, Vitale, \&
  Loeb}]{Christian:2018vsi}
Christian, P., Vitale, S., \& Loeb, A. 2018, Phys. Rev. D, 98, 103022,
  \dodoi{10.1103/PhysRevD.98.103022}

\bibitem[{Colleoni {et~al.}(2021)Colleoni, Mateu-Lucena, Estell\'es,
  Garc\'\i{}a-Quir\'os, Keitel, Pratten, Ramos-Buades, \&
  Husa}]{Colleoni:2020tgc}
Colleoni, M., Mateu-Lucena, M., Estell\'es, H., {et~al.} 2021, Phys. Rev. D,
  103, 024029, \dodoi{10.1103/PhysRevD.103.024029}

\bibitem[{Collett \& Bacon(2017)}]{Collett:2016dey}
Collett, T.~E., \& Bacon, D. 2017, Phys. Rev. Lett., 118, 091101,
  \dodoi{10.1103/PhysRevLett.118.091101}

\bibitem[{Cornish {et~al.}(2011)Cornish, Sampson, Yunes, \&
  Pretorius}]{Cornish:2011ys}
Cornish, N., Sampson, L., Yunes, N., \& Pretorius, F. 2011, \prd, 84, 062003,
  \dodoi{10.1103/PhysRevD.84.062003}

\bibitem[{Cornish {et~al.}(2021)Cornish, Littenberg, B\'ecsy, Chatziioannou,
  Clark, Ghonge, \& Millhouse}]{PhysRevD.103.044006}
Cornish, N.~J., Littenberg, T.~B., B\'ecsy, B., {et~al.} 2021, Phys. Rev. D,
  103, 044006, \dodoi{10.1103/PhysRevD.103.044006}

\bibitem[{Cremonese {et~al.}(2021)Cremonese, Ezquiaga, \&
  Salzano}]{Cremonese:2021puh}
Cremonese, P., Ezquiaga, J.~M., \& Salzano, V. 2021, Phys. Rev. D, 104, 023503,
  \dodoi{10.1103/PhysRevD.104.023503}

\bibitem[{Dai {et~al.}(2018)Dai, Li, Zackay, Mao, \& Lu}]{Dai:2018enj}
Dai, L., Li, S.-S., Zackay, B., Mao, S., \& Lu, Y. 2018, Phys. Rev. D, 98,
  104029, \dodoi{10.1103/PhysRevD.98.104029}

\bibitem[{Dai \& Venumadhav(2017)}]{Dai:2017huk}
Dai, L., \& Venumadhav, T. 2017.
\newblock \doarXiv{1702.04724}

\bibitem[{Dai {et~al.}(2020)Dai, Zackay, Venumadhav, Roulet, \&
  Zaldarriaga}]{Dai:2020tpj}
Dai, L., Zackay, B., Venumadhav, T., Roulet, J., \& Zaldarriaga, M. 2020.
\newblock \doarXiv{2007.12709}

\bibitem[{Dal~Canton {et~al.}(2014)}]{Canton:2014ena}
Dal~Canton, T., {et~al.} 2014, Phys. Rev. D, 90, 082004,
  \dodoi{10.1103/PhysRevD.90.082004}

\bibitem[{Damour \& Nagar(2016)}]{Damour:2016bks}
Damour, T., \& Nagar, A. 2016, Lect. Notes Phys., 905, 273,
  \dodoi{10.1007/978-3-319-19416-5_7}

\bibitem[{Davis {et~al.}(2022)Davis, Littenberg, Romero-Shaw, Millhouse,
  McIver, Di~Renzo, \& Ashton}]{Davis_glitchSub22}
Davis, D., Littenberg, T.~B., Romero-Shaw, I.~M., {et~al.} 2022,
  \dodoi{10.48550/ARXIV.2207.03429}

\bibitem[{Davis {et~al.}(2019)Davis, Massinger, Lundgren, Driggers, Urban, \&
  Nuttall}]{Davis:2018yrz}
Davis, D., Massinger, T.~J., Lundgren, A.~P., {et~al.} 2019, Class. Quant.
  Grav., 36, 055011, \dodoi{10.1088/1361-6382/ab01c5}

\bibitem[{Davis {et~al.}(2021{\natexlab{a}})}]{Davis2021}
Davis, D., {et~al.} 2021{\natexlab{a}}, Class. Quant. Grav., 38, 135014,
  \dodoi{10.1088/1361-6382/abfd85}

\bibitem[{Davis {et~al.}(2021{\natexlab{b}})}]{Davis:2021ecd}
---. 2021{\natexlab{b}}, Class. Quant. Grav., 38, 135014,
  \dodoi{10.1088/1361-6382/abfd85}

\bibitem[{Deguchi \& Watson(1986)}]{Deguchi:1986zz}
Deguchi, S., \& Watson, W.~D. 1986, Phys. Rev. D, 34, 1708,
  \dodoi{10.1103/PhysRevD.34.1708}

\bibitem[{Deng {et~al.}(2009)Deng, Dong, Socher, Li, Li, \&
  Fei-Fei}]{Deng:2009ML}
Deng, J., Dong, W., Socher, R., {et~al.} 2009, in 2009 IEEE Conference on
  Computer Vision and Pattern Recognition, 248--255,
  \dodoi{10.1109/CVPR.2009.5206848}

\bibitem[{Diego {et~al.}(2019)Diego, Hannuksela, Kelly, Broadhurst, Kim, Li,
  Smoot, \& Pagano}]{Diego:2019lcd}
Diego, J., Hannuksela, O., Kelly, P., {et~al.} 2019, Astron. Astrophys., 627,
  A130, \dodoi{10.1051/0004-6361/201935490}

\bibitem[{Diego(2020)}]{Diego:2019rzc}
Diego, J.~M. 2020, Phys. Rev. D, 101, 123512,
  \dodoi{10.1103/PhysRevD.101.123512}

\bibitem[{Dominik {et~al.}(2013)Dominik, Belczynski, Fryer, Holz, Berti, Bulik,
  Mandel, \& O'Shaughnessy}]{Dominik:2013tma}
Dominik, M., Belczynski, K., Fryer, C., {et~al.} 2013, Astrophys. J., 779, 72,
  \dodoi{10.1088/0004-637X/779/1/72}

\bibitem[{{Dominik} {et~al.}(2013){Dominik}, {Belczynski}, {Fryer}, {Holz},
  {Berti}, {Bulik}, {Mandel}, \& {O'Shaughnessy}}]{Dominik:2013}
{Dominik}, M., {Belczynski}, K., {Fryer}, C., {et~al.} 2013, \apj, 779, 72,
  \dodoi{10.1088/0004-637X/779/1/72}

\bibitem[{Driggers {et~al.}(2019)}]{Driggers:2018gii}
Driggers, J.~C., {et~al.} 2019, Phys. Rev. D, 99, 042001,
  \dodoi{10.1103/PhysRevD.99.042001}

\bibitem[{Eldridge {et~al.}(2019)Eldridge, Stanway, \& Tang}]{Eldridge:2018nop}
Eldridge, J., Stanway, E., \& Tang, P.~N. 2019, Mon. Not. Roy. Astron. Soc.,
  482, 870, \dodoi{10.1093/mnras/sty2714}

\bibitem[{Ezquiaga {et~al.}(2021)Ezquiaga, Holz, Hu, Lagos, \&
  Wald}]{Ezquiaga:2020gdt}
Ezquiaga, J.~M., Holz, D.~E., Hu, W., Lagos, M., \& Wald, R.~M. 2021, Phys.
  Rev. D, 103, 6, \dodoi{10.1103/PhysRevD.103.064047}

\bibitem[{Ezquiaga {et~al.}(2022)Ezquiaga, Hu, Lagos, Lin, \&
  Xu}]{Ezquiaga:2022nak}
Ezquiaga, J.~M., Hu, W., Lagos, M., Lin, M.-X., \& Xu, F. 2022.
\newblock \doarXiv{2203.13252}

\bibitem[{Ezquiaga \& Zumalac\'arregui(2020)}]{Ezquiaga:2020dao}
Ezquiaga, J.~M., \& Zumalac\'arregui, M. 2020, Phys. Rev. D, 102, 124048,
  \dodoi{10.1103/PhysRevD.102.124048}

\bibitem[{Fan {et~al.}(2017)Fan, Liao, Biesiada, Piorkowska-Kurpas, \&
  Zhu}]{Fan:2016swi}
Fan, X.-L., Liao, K., Biesiada, M., Piorkowska-Kurpas, A., \& Zhu, Z.-H. 2017,
  Phys. Rev. Lett., 118, 091102, \dodoi{10.1103/PhysRevLett.118.091102}

\bibitem[{Farr {et~al.}(2015)Farr, Gair, Mandel, \& Cutler}]{Farr:2013yna}
Farr, W.~M., Gair, J.~R., Mandel, I., \& Cutler, C. 2015, Phys. Rev. D, 91,
  023005, \dodoi{10.1103/PhysRevD.91.023005}

\bibitem[{{Finn} \& {Chernoff}(1993)}]{FinnChernoff:1993}
{Finn}, L.~S., \& {Chernoff}, D.~F. 1993, \prd, 47, 2198,
  \dodoi{10.1103/PhysRevD.47.2198}

\bibitem[{Fiori {et~al.}(2020)}]{Fiori2020}
Fiori, I., {et~al.} 2020, Galaxies, 8, 82, \dodoi{10.3390/galaxies8040082}

\bibitem[{Goyal {et~al.}(2021{\natexlab{a}})Goyal, D., Kapadia, \&
  Ajith}]{Goyal:2021hxv}
Goyal, S., D., H., Kapadia, S.~J., \& Ajith, P. 2021{\natexlab{a}}.
\newblock \doarXiv{2106.12466}

\bibitem[{Goyal {et~al.}(2021{\natexlab{b}})Goyal, Haris, Mehta, \&
  Ajith}]{Goyal:2020bkm}
Goyal, S., Haris, K., Mehta, A.~K., \& Ajith, P. 2021{\natexlab{b}}, Phys. Rev.
  D, 103, 024038, \dodoi{10.1103/PhysRevD.103.024038}

\bibitem[{{GWOSC}(2021)}]{gwosc:gwtc3}
{GWOSC}. 2021, {GWTC-3 Data Release}, \dodoi{10.7935/b024-1886}

\bibitem[{Hanna {et~al.}(2020)}]{Hanna:2019ezx}
Hanna, C., {et~al.} 2020, Phys. Rev. D, 101, 022003,
  \dodoi{10.1103/PhysRevD.101.022003}

\bibitem[{Hannam {et~al.}(2021)Hannam, Hoy, Thompson, Fairhurst, \&
  Raymond}]{Hannam:2021pit}
Hannam, M., Hoy, C., Thompson, J.~E., Fairhurst, S., \& Raymond, V. 2021.
\newblock \doarXiv{2112.11300}

\bibitem[{Hannuksela {et~al.}(2020)Hannuksela, Collett, \c{C}al\i{}\c{s}kan, \&
  Li}]{Hannuksela:2020xor}
Hannuksela, O.~A., Collett, T.~E., \c{C}al\i{}\c{s}kan, M., \& Li, T. G.~F.
  2020, Mon. Not. Roy. Astron. Soc., 498, 3395, \dodoi{10.1093/mnras/staa2577}

\bibitem[{Haris {et~al.}(2018)Haris, Mehta, Kumar, Venumadhav, \&
  Ajith}]{Haris:2018vmn}
Haris, K., Mehta, A.~K., Kumar, S., Venumadhav, T., \& Ajith, P. 2018.
\newblock \doarXiv{1807.07062}

\bibitem[{Harris {et~al.}(2020)}]{Harris:2020xlr}
Harris, C.~R., {et~al.} 2020, Nature, 585, 357,
  \dodoi{10.1038/s41586-020-2649-2}

\bibitem[{{Huang} {et~al.}(2016){Huang}, {Liu}, {van der Maaten}, \&
  {Weinberger}}]{Huang:2016ML}
{Huang}, G., {Liu}, Z., {van der Maaten}, L., \& {Weinberger}, K.~Q. 2016,
  arXiv e-prints, arXiv:1608.06993.
\newblock \doarXiv{1608.06993}

\bibitem[{{Hunter}(2007)}]{matplotlib}
{Hunter}, J.~D. 2007, Computing in Science Engineering, 9, 90,
  \dodoi{10.1109/MCSE.2007.55}

\bibitem[{Janquart {et~al.}(2021)Janquart, Hannuksela, K., \& Van
  Den~Broeck}]{Janquart:2021qov}
Janquart, J., Hannuksela, O.~A., K., H., \& Van Den~Broeck, C. 2021,
  \dodoi{10.1093/mnras/stab1991}

\bibitem[{Jung \& Shin(2019)}]{Jung:2017flg}
Jung, S., \& Shin, C.~S. 2019, Phys. Rev. Lett., 122, 041103,
  \dodoi{10.1103/PhysRevLett.122.041103}

\bibitem[{Kapadia {et~al.}(2020)}]{Kapadia:2019uut}
Kapadia, S.~J., {et~al.} 2020, Class. Quant. Grav., 37, 045007,
  \dodoi{10.1088/1361-6382/ab5f2d}

\bibitem[{Karki {et~al.}(2016)}]{Karki:2016pht}
Karki, S., {et~al.} 2016, Rev. Sci. Instrum., 87, 114503,
  \dodoi{10.1063/1.4967303}

\bibitem[{Khan {et~al.}(2016)Khan, Husa, Hannam, Ohme, P\"urrer, Forteza, \&
  Boh\'e}]{Khan_2016IMRPhenomD}
Khan, S., Husa, S., Hannam, M., {et~al.} 2016, Phys. Rev. D, 93, 044007,
  \dodoi{10.1103/PhysRevD.93.044007}

\bibitem[{Khan {et~al.}(2020)Khan, Ohme, Chatziioannou, \&
  Hannam}]{Khan:2019kot}
Khan, S., Ohme, F., Chatziioannou, K., \& Hannam, M. 2020, Phys. Rev. D, 101,
  024056, \dodoi{10.1103/PhysRevD.101.024056}

\bibitem[{Klimenko {et~al.}(2005)Klimenko, Mohanty, Rakhmanov, \&
  Mitselmakher}]{Klimenko:2005xv}
Klimenko, S., Mohanty, S., Rakhmanov, M., \& Mitselmakher, G. 2005, Phys.\
  Rev.\ D, 72, 122002, \dodoi{10.1103/PhysRevD.72.122002}

\bibitem[{Klimenko {et~al.}(2006)Klimenko, Mohanty, Rakhmanov, \&
  Mitselmakher}]{Klimenko:2006rh}
---. 2006, J.\ Phys.\ Conf.\ Ser., 32, 12, \dodoi{10.1088/1742-6596/32/1/003}

\bibitem[{Klimenko {et~al.}(2004)Klimenko, Yakushin, Rakhmanov, \&
  Mitselmakher}]{Klimenko:2004qh}
Klimenko, S., Yakushin, I., Rakhmanov, M., \& Mitselmakher, G. 2004, Class.
  Quant. Grav., 21, S1685, \dodoi{10.1088/0264-9381/21/20/011}

\bibitem[{Klimenko {et~al.}(2011)Klimenko, Vedovato, Drago, Mazzolo,
  Mitselmakher, Pankow, Prodi, Re, Salemi, \& Yakushin}]{Klimenko:2011hz}
Klimenko, S., Vedovato, G., Drago, M., {et~al.} 2011, Phys. Rev. D, 83, 102001,
  \dodoi{10.1103/PhysRevD.83.102001}

\bibitem[{Klimenko {et~al.}(2016)}]{Klimenko:2015ypf}
Klimenko, S., {et~al.} 2016, Phys. Rev. D, 93, 042004,
  \dodoi{10.1103/PhysRevD.93.042004}

\bibitem[{Lai {et~al.}(2018)Lai, Hannuksela, Herrera-Mart{\'\i}n, Diego,
  Broadhurst, \& Li}]{Lai:2018rto}
Lai, K.-H., Hannuksela, O.~A., Herrera-Mart{\'\i}n, A., {et~al.} 2018, Phys.
  Rev. D, 98, 083005, \dodoi{10.1103/PhysRevD.98.083005}

\bibitem[{Lange {et~al.}(2017)Lange, O'Shaughnessy, Boyle, Calder\'on~Bustillo,
  Campanelli, Chu, Clark, Demos, Fong, Healy, Hemberger, Hinder, Jani,
  Khamesra, Kidder, Kumar, Laguna, Lousto, Lovelace, Ossokine, Pfeiffer,
  Scheel, Shoemaker, Szilagyi, Teukolsky, \& Zlochower}]{PhysRevD.96.104041}
Lange, J., O'Shaughnessy, R., Boyle, M., {et~al.} 2017, Phys. Rev. D, 96,
  104041, \dodoi{10.1103/PhysRevD.96.104041}

\bibitem[{Li {et~al.}(2019{\natexlab{a}})Li, Lo, Sachdev, Chan, Lin, Li, \&
  Weinstein}]{Li:2019osa}
Li, A.~K., Lo, R.~K., Sachdev, S., {et~al.} 2019{\natexlab{a}}.
\newblock \doarXiv{1904.06020}

\bibitem[{Li {et~al.}(2018)Li, Mao, Zhao, \& Lu}]{Li:2018prc}
Li, S.-S., Mao, S., Zhao, Y., \& Lu, Y. 2018, Mon. Not. Roy. Astron. Soc., 476,
  2220, \dodoi{10.1093/mnras/sty411}

\bibitem[{Li {et~al.}(2019{\natexlab{b}})Li, Fan, \& Gou}]{Li:2019rns}
Li, Y., Fan, X., \& Gou, L. 2019{\natexlab{b}}, Astrophys. J., 873, 37,
  \dodoi{10.3847/1538-4357/ab037e}

\bibitem[{Liao {et~al.}(2017)Liao, Fan, Ding, Biesiada, \& Zhu}]{Liao:2017ioi}
Liao, K., Fan, X.-L., Ding, X.-H., Biesiada, M., \& Zhu, Z.-H. 2017, Nature
  Commun., 8, 1148, \dodoi{10.1038/s41467-017-01152-9}

\bibitem[{{LIGO Scientific Collaboration and Virgo
  Collaboration}(2018)}]{lalsuite}
{LIGO Scientific Collaboration and Virgo Collaboration}. 2018, {LALSuite
  software}, \dodoi{10.7935/GT1W-FZ16}

\bibitem[{Lo \& Maga\~na Hernandez(2021)}]{Lo:2021nae}
Lo, R. K.~L., \& Maga\~na Hernandez, I. 2021.
\newblock \doarXiv{2104.09339}

\bibitem[{Madau \& Dickinson(2014)}]{Madau:2014bja}
Madau, P., \& Dickinson, M. 2014, Ann. Rev. Astron. Astrophys., 52, 415,
  \dodoi{10.1146/annurev-astro-081811-125615}

\bibitem[{McIsaac {et~al.}(2020)McIsaac, Keitel, Collett, Harry, Mozzon, Edy,
  \& Bacon}]{McIsaac:2019use}
McIsaac, C., Keitel, D., Collett, T., {et~al.} 2020, Phys. Rev. D, 102, 084031,
  \dodoi{10.1103/PhysRevD.102.084031}

\bibitem[{{Messick} {et~al.}(2016){Messick}, {Blackburn}, {Brady}, {Brockill},
  {Cannon}, {Caudill}, {Chamberlin}, {Creighton}, {Everett}, {Hanna}, {Lang},
  {Li}, {Meacher}, {Pankow}, {Privitera}, {Qi}, {Sachdev}, {Sadeghian},
  {Sathyaprakash}, {Singer}, {Thomas}, {Wade}, {Wade}, \&
  {Weinstein}}]{gstlal-methods}
{Messick}, C., {Blackburn}, K., {Brady}, P., {et~al.} 2016.
\newblock \doarXiv{1604.04324}

\bibitem[{Messick {et~al.}(2017)}]{Messick:2016aqy}
Messick, C., {et~al.} 2017, Phys. Rev. D, 95, 042001,
  \dodoi{10.1103/PhysRevD.95.042001}

\bibitem[{Mishra {et~al.}(2021)Mishra, Meena, More, Bose, \&
  Bagla}]{Mishra:2021xzz}
Mishra, A., Meena, A.~K., More, A., Bose, S., \& Bagla, J.~S. 2021, Mon. Not.
  Roy. Astron. Soc., 508, 4869, \dodoi{10.1093/mnras/stab2875}

\bibitem[{Mukherjee {et~al.}(2021{\natexlab{a}})Mukherjee, Broadhurst, Diego,
  Silk, \& Smoot}]{Mukherjee:2021qam}
Mukherjee, S., Broadhurst, T., Diego, J.~M., Silk, J., \& Smoot, G.~F.
  2021{\natexlab{a}}.
\newblock \doarXiv{2106.00392}

\bibitem[{Mukherjee {et~al.}(2021{\natexlab{b}})Mukherjee, Broadhurst, Diego,
  Silk, \& Smoot}]{Mukherjee:2020tvr}
---. 2021{\natexlab{b}}, Mon. Not. Roy. Astron. Soc., 501, 2451,
  \dodoi{10.1093/mnras/staa3813}

\bibitem[{Nakamura(1998)}]{Nakamura:1997sw}
Nakamura, T.~T. 1998, Phys. Rev. Lett., 80, 1138,
  \dodoi{10.1103/PhysRevLett.80.1138}

\bibitem[{Ng {et~al.}(2018)Ng, Wong, Broadhurst, \& Li}]{Ng:2017yiu}
Ng, K.~K., Wong, K.~W., Broadhurst, T., \& Li, T.~G. 2018, Phys. Rev. D, 97,
  023012, \dodoi{10.1103/PhysRevD.97.023012}

\bibitem[{Nguyen {et~al.}(2021)}]{nguyen2021environmental}
Nguyen, P., {et~al.} 2021, Class. Quant. Grav., 38, 145001,
  \dodoi{10.1088/1361-6382/ac011a}

\bibitem[{Nitz {et~al.}(2017)Nitz, Dent, Dal~Canton, Fairhurst, \&
  Brown}]{Nitz:2017svb}
Nitz, A.~H., Dent, T., Dal~Canton, T., Fairhurst, S., \& Brown, D.~A. 2017,
  Astrophys. J., 849, 118, \dodoi{10.3847/1538-4357/aa8f50}

\bibitem[{Oguri(2018)}]{Oguri:2018muv}
Oguri, M. 2018, Mon. Not. Roy. Astron. Soc., 480, 3842,
  \dodoi{10.1093/mnras/sty2145}

\bibitem[{{Oguri} \& {Marshall}(2010)}]{Oguri:2010a}
{Oguri}, M., \& {Marshall}, P.~J. 2010, Mon. Not. Roy. Astron. Soc., 405, 2579,
  \dodoi{10.1111/j.1365-2966.2010.16639.x}

\bibitem[{Oguri \& Takahashi(2020)}]{Oguri:2020ldf}
Oguri, M., \& Takahashi, R. 2020, Astrophys. J., 901, 58,
  \dodoi{10.3847/1538-4357/abafab}

\bibitem[{Ohanian(1974)}]{Ohanian:1974ys}
Ohanian, H. 1974, Int. J. Theor. Phys., 9, 425, \dodoi{10.1007/BF01810927}

\bibitem[{Ossokine {et~al.}(2020)}]{Ossokine:2020kjp}
Ossokine, S., {et~al.} 2020, Phys. Rev. D, 102, 044055,
  \dodoi{10.1103/PhysRevD.102.044055}

\bibitem[{Pagano {et~al.}(2020)Pagano, Hannuksela, \& Li}]{Pagano:2020rwj}
Pagano, G., Hannuksela, O.~A., \& Li, T. G.~F. 2020, Astron. Astrophys., 643,
  A167, \dodoi{10.1051/0004-6361/202038730}

\bibitem[{Palenzuela(2020)}]{Palenzuela:2020tga}
Palenzuela, C. 2020, Front. Astron. Space Sci., 7, 58,
  \dodoi{10.3389/fspas.2020.00058}

\bibitem[{Pankow {et~al.}(2015)Pankow, Brady, Ochsner, \&
  O'Shaughnessy}]{PhysRevD.92.023002}
Pankow, C., Brady, P., Ochsner, E., \& O'Shaughnessy, R. 2015, Phys. Rev. D,
  92, 023002, \dodoi{10.1103/PhysRevD.92.023002}

\bibitem[{Payne {et~al.}(2022)Payne, Hourihane, Golomb, Udall, Davis, \&
  Chatziioannou}]{Payne:2022spz}
Payne, E., Hourihane, S., Golomb, J., {et~al.} 2022.
\newblock \doarXiv{2206.11932}

\bibitem[{{Perez} \& {Granger}(2007)}]{ipython}
{Perez}, F., \& {Granger}, B.~E. 2007, Computing in Science Engineering, 9, 21,
  \dodoi{10.1109/MCSE.2007.53}

\bibitem[{Pratten {et~al.}(2021)}]{Pratten:2020ceb}
Pratten, G., {et~al.} 2021, Phys. Rev. D, 103, 104056,
  \dodoi{10.1103/PhysRevD.103.104056}

\bibitem[{Price-Whelan {et~al.}(2018)}]{astropy:2018}
Price-Whelan, A., {et~al.} 2018, Astron. J., 156, 123,
  \dodoi{10.3847/1538-3881/aabc4f}

\bibitem[{Robertson {et~al.}(2020)Robertson, Smith, Massey, Eke, Jauzac,
  Bianconi, \& Ryczanowski}]{Robertson:2020mfh}
Robertson, A., Smith, G.~P., Massey, R., {et~al.} 2020,
  \dodoi{10.1093/mnras/staa1429}

\bibitem[{Robitaille {et~al.}(2013)}]{astropy:2013}
Robitaille, T.~P., {et~al.} 2013, Astron. Astrophys., 558, A33,
  \dodoi{10.1051/0004-6361/201322068}

\bibitem[{Romero-Shaw {et~al.}(2020)}]{Romero-Shaw:2020owr}
Romero-Shaw, I.~M., {et~al.} 2020, Mon. Not. Roy. Astron. Soc., 499, 3295,
  \dodoi{10.1093/mnras/staa2850}

\bibitem[{Ryczanowski {et~al.}(2020)Ryczanowski, Smith, Bianconi, Massey,
  Robertson, \& Jauzac}]{Ryczanowski:2020mlt}
Ryczanowski, D., Smith, G.~P., Bianconi, M., {et~al.} 2020, Mon. Not. Roy.
  Astron. Soc., 495, 1666, \dodoi{10.1093/mnras/staa1274}

\bibitem[{Sachdev {et~al.}(2019)}]{Sachdev:2019vvd}
Sachdev, S., {et~al.} 2019.
\newblock \doarXiv{1901.08580}

\bibitem[{Schmidt(2020)}]{Schmidt:2020ekt}
Schmidt, P. 2020, Front. Astron. Space Sci., 7, 28,
  \dodoi{10.3389/fspas.2020.00028}

\bibitem[{{Schneider} {et~al.}(1992){Schneider}, {Ehlers}, \&
  {Falco}}]{1992grle.book.....S}
{Schneider}, P., {Ehlers}, J., \& {Falco}, E.~E. 1992, {Gravitational Lenses},
  112, \dodoi{10.1007/978-3-662-03758-4}

\bibitem[{Sereno {et~al.}(2011)Sereno, Jetzer, Sesana, \&
  Volonteri}]{Sereno:2011ty}
Sereno, M., Jetzer, P., Sesana, A., \& Volonteri, M. 2011, Mon. Not. Roy.
  Astron. Soc., 415, 2773, \dodoi{10.1111/j.1365-2966.2011.18895.x}

\bibitem[{{Singer}(2019)}]{ligoskymap}
{Singer}, L. 2019, ligo.skymap \url{https://lscsoft.docs.ligo.org/ligo.skymap/}

\bibitem[{Singer \& Price(2016)}]{Singer:2015ema}
Singer, L.~P., \& Price, L.~R. 2016, Phys. Rev., D93, 024013,
  \dodoi{10.1103/PhysRevD.93.024013}

\bibitem[{Smith {et~al.}(2017)}]{Smith:2018gle}
Smith, G., {et~al.} 2017, IAU Symp., 338, 98, \dodoi{10.1017/S1743921318003757}

\bibitem[{Smith {et~al.}(2018)Smith, Jauzac, Veitch, Farr, Massey, \&
  Richard}]{Smith:2017mqu}
Smith, G.~P., Jauzac, M., Veitch, J., {et~al.} 2018, Mon. Not. Roy. Astron.
  Soc., 475, 3823, \dodoi{10.1093/mnras/sty031}

\bibitem[{Smith {et~al.}(2019)Smith, Robertson, Bianconi, \&
  Jauzac}]{Smith:2019dis}
Smith, G.~P., Robertson, A., Bianconi, M., \& Jauzac, M. 2019.
\newblock \doarXiv{1902.05140}

\bibitem[{Smith {et~al.}(2020)Smith, Ashton, Vajpeyi, \&
  Talbot}]{Smith:2019ucc}
Smith, R. J.~E., Ashton, G., Vajpeyi, A., \& Talbot, C. 2020, Mon. Not. Roy.
  Astron. Soc., 498, 4492, \dodoi{10.1093/mnras/staa2483}

\bibitem[{Speagle(2020)}]{10.1093/mnras/staa278}
Speagle, J.~S. 2020, Monthly Notices of the Royal Astronomical Society, 493,
  3132, \dodoi{10.1093/mnras/staa278}

\bibitem[{Sun {et~al.}(2020)Sun, Goetz, Kissel, Betzwieser, Karki, Viets, Wade,
  Bhattacharjee, Bossilkov, Covas, Datrier, Gray, Kandhasamy, Lecoeuche,
  Mendell, Mistry, Payne, Savage, Weinstein, Aston, Buikema, Cahillane,
  Driggers, Dwyer, Kumar, \& Urban}]{Sun_2020}
Sun, L., Goetz, E., Kissel, J.~S., {et~al.} 2020, Classical and Quantum
  Gravity, 37, 225008, \dodoi{10.1088/1361-6382/abb14e}

\bibitem[{Sun {et~al.}(2021)}]{Sun2021}
Sun, L., {et~al.} 2021.
\newblock \doarXiv{2107.00129}

\bibitem[{Takahashi \& Nakamura(2003)}]{Takahashi:2003ix}
Takahashi, R., \& Nakamura, T. 2003, Astrophys. J., 595, 1039,
  \dodoi{10.1086/377430}

\bibitem[{Talbot {et~al.}(2019)Talbot, Smith, Thrane, \&
  Poole}]{Talbot:2019okv}
Talbot, C., Smith, R., Thrane, E., \& Poole, G.~B. 2019, Phys. Rev. D, 100,
  043030, \dodoi{10.1103/PhysRevD.100.043030}

\bibitem[{Thorne(1982)}]{Thorne:1982cv}
Thorne, K. 1982, in {Les Houches Summer School on Gravitational Radiation},
  1--57

\bibitem[{Tinker {et~al.}(2008)Tinker, Kravtsov, Klypin, Abazajian, Warren,
  Yepes, Gottlober, \& Holz}]{Tinker:2008ff}
Tinker, J.~L., Kravtsov, A.~V., Klypin, A., {et~al.} 2008, Astrophys. J., 688,
  709, \dodoi{10.1086/591439}

\bibitem[{Urrutia \& Vaskonen(2021)}]{Urrutia:2021qak}
Urrutia, J., \& Vaskonen, V. 2021.
\newblock \doarXiv{2109.03213}

\bibitem[{Usman {et~al.}(2016)}]{Usman:2015kfa}
Usman, S.~A., {et~al.} 2016, Class. Quant. Grav., 33, 215004,
  \dodoi{10.1088/0264-9381/33/21/215004}

\bibitem[{Vajente {et~al.}(2020)Vajente, Huang, Isi, Driggers, Kissel,
  Szczepanczyk, \& Vitale}]{Vajente:2019ycy}
Vajente, G., Huang, Y., Isi, M., {et~al.} 2020, Phys.\ Rev.\ D, 101, 042003,
  \dodoi{10.1103/PhysRevD.101.042003}

\bibitem[{Vallisneri(2012)}]{Vallisneri:2012qq}
Vallisneri, M. 2012, Phys. Rev. D, 86, 082001,
  \dodoi{10.1103/PhysRevD.86.082001}

\bibitem[{Viets {et~al.}(2018)}]{Viets:2017yvy}
Viets, A., {et~al.} 2018, Class. Quant. Grav., 35, 095015,
  \dodoi{10.1088/1361-6382/aab658}

\bibitem[{{Virtanen} {et~al.}(2020){Virtanen}, {Gommers}, {Oliphant},
  {Haberland}, {Reddy}, {Cournapeau}, {Burovski}, {Peterson}, {Weckesser},
  {Bright}, {van der Walt}, {Brett}, {Wilson}, {Jarrod Millman}, {Mayorov},
  {Nelson}, {Jones}, {Kern}, {Larson}, {Carey}, {Polat}, {Feng}, {Moore}, {Vand
  erPlas}, {Laxalde}, {Perktold}, {Cimrman}, {Henriksen}, {Quintero}, {Harris},
  {Archibald}, {Ribeiro}, {Pedregosa}, {van Mulbregt}, \&
  {Contributors}}]{2020SciPy-NMeth}
{Virtanen}, P., {Gommers}, R., {Oliphant}, T.~E., {et~al.} 2020, Nature
  Methods, 17, 261, \dodoi{https://doi.org/10.1038/s41592-019-0686-2}

\bibitem[{Wang {et~al.}(1996)Wang, Stebbins, \& Turner}]{Wang:1996as}
Wang, Y., Stebbins, A., \& Turner, E.~L. 1996, Phys. Rev. Lett., 77, 2875,
  \dodoi{10.1103/PhysRevLett.77.2875}

\bibitem[{Waskom {et~al.}(2020)Waskom, Botvinnik, Ostblom, Gelbart, Lukauskas,
  Hobson, Gemperline, Augspurger, Halchenko, Cole, Warmenhoven, de~Ruiter, Pye,
  Hoyer, Vanderplas, Villalba, Kunter, Quintero, Bachant, Martin, Meyer, Swain,
  Miles, Brunner, O'Kane, Yarkoni, Williams, Evans, Fitzgerald, \&
  Brian}]{seaborn}
Waskom, M., Botvinnik, O., Ostblom, J., {et~al.} 2020, mwaskom/seaborn: v0.10.1
  (April 2020), v0.10.1,  Zenodo, \dodoi{10.5281/zenodo.3767070}

\bibitem[{Wempe {et~al.}(2022)Wempe, Koopmans, Wierda, Hannuksela, \&
  Broeck}]{Wempe:2022zlk}
Wempe, E., Koopmans, L. V.~E., Wierda, A. R. A.~C., Hannuksela, O.~A., \&
  Broeck, C. v.~d. 2022.
\newblock \doarXiv{2204.08732}

\bibitem[{Wierda {et~al.}(2021)Wierda, Wempe, Hannuksela, Koopmans, \& Van
  Den~Broeck}]{Wierda:2021upe}
Wierda, A. R. A.~C., Wempe, E., Hannuksela, O.~A., Koopmans, L. V.~E., \& Van
  Den~Broeck, C. 2021.
\newblock \doarXiv{2106.06303}

\bibitem[{Wong {et~al.}(2021)Wong, Chan, Wong, Lo, \& Li}]{Wong:2021lxf}
Wong, H. W.~Y., Chan, L. W.~L., Wong, I. C.~F., Lo, R. K.~L., \& Li, T. G.~F.
  2021.
\newblock \doarXiv{2112.05932}

\bibitem[{Wysocki {et~al.}(2019)Wysocki, O'Shaughnessy, Lange, \&
  Fang}]{PhysRevD.99.084026}
Wysocki, D., O'Shaughnessy, R., Lange, J., \& Fang, Y.-L.~L. 2019, Phys. Rev.
  D, 99, 084026, \dodoi{10.1103/PhysRevD.99.084026}

\bibitem[{Xu {et~al.}(2021)Xu, Ezquiaga, \& Holz}]{Xu:2021bfn}
Xu, F., Ezquiaga, J.~M., \& Holz, D.~E. 2021.
\newblock \doarXiv{2105.14390}

\bibitem[{Yeung {et~al.}(2021)Yeung, Cheung, Gais, Hannuksela, \&
  Li}]{Yeung:2021roe}
Yeung, S. M.~C., Cheung, M. H.~Y., Gais, J. A.~J., Hannuksela, O.~A., \& Li, T.
  G.~F. 2021.
\newblock \doarXiv{2112.07635}

\bibitem[{Zevin {et~al.}(2021)Zevin, Bavera, Berry, Kalogera, Fragos, Marchant,
  Rodriguez, Antonini, Holz, \& Pankow}]{Zevin:2020gbd}
Zevin, M., Bavera, S.~S., Berry, C. P.~L., {et~al.} 2021, Astrophys. J., 910,
  152, \dodoi{10.3847/1538-4357/abe40e}

\end{thebibliography}

\clearpage

\iftoggle{endauthorlist}{
 \let\author\myauthor
 \let\affiliation\myaffiliation
 \let\maketitle\mymaketitle
 
 \pacs{}
 \maketitle
}

\end{document}